\documentclass{aa}
\usepackage{graphics,psfig}
\begin{document}

   \thesaurus{11     
              (11.05.2; 
11.19.2; 
11.19.7; 
11.19.6)} 

\title{Box- and peanut-shaped bulges
\thanks{
Based on observations collected at ESO/La Silla (61.A-0143), DSAZ/Calar Alto,
and TIRGO/Gornergrat.}
}

\subtitle{II. NIR observations}

\author{R. L\"utticke \and R.-J. Dettmar \and M. Pohlen}

\institute{Astronomisches Institut, Ruhr-Universit\"at Bochum,
D-44780 Bochum, Germany\\
email: luett@astro.ruhr-uni-bochum.de
}

\date{Received 19 July 2000 / Accepted 28 August 2000}

\titlerunning{Box- and peanut-shaped bulges. II}
\maketitle

\begin{abstract}

We have observed 60 edge-on galaxies  in the NIR in
order to study the stellar distribution in galaxies with box/peanut-shaped
bulges.
The much smaller amount of dust extinction at these wavelengths
allows us to identify in almost all target galaxies with
box/peanut-shaped bulges an additional thin, central component in
cuts parallel to the major axis.
This structure can be identified
with a bar. \\
The length of this structure scaled by the length of the bulge
correlates with the morphologically classified
shape of the bulge. This newly established correlation is therefore 
mainly interpreted
as the projection of the bar at different aspect angles. 
Galaxies with peanut bulges have a bar seen nearly edge-on and
the ratio of bar length to thickness, $14 \pm 4$,
can be directly measured for the first time.
In addition, the correlation of the boxiness of bulges
with the bar strength indicates that the bar characteristic could partly
explain differences in the bulge shape. 
Furthermore, a new size relation between the box/peanut
structure and the central bulge is found. 
Our observations are discussed in comparison to a N-body
simulation for barred galaxies (Pfenniger \& Friedli \cite{pfe}).
We conclude that the inner region of barred disk galaxies are build up
by three distinct components:
the spheroidal bulge, a thin bar, and a b/p structure most likely
representing the thick part of the bar.


\end{abstract}

\section{Introduction}
Our statistical study of
$\sim$\,1350 edge-on galaxies
derived from the RC3 (\emph {Third Reference Catalogue of Bright Galaxies},
de Vaucouleurs et al. \cite{rc3})
(L\"utticke et al. \cite{lue2000a}, hereafter Paper I)
has revealed that 45\,\% of all bulges (S0-Sd galaxies)
are box- and peanut-shaped (b/p).
This high frequency must be considered in a theory
of bulge formation and especially in
scenarios explaining the formation process
of b/p bulges.

Currently discussed theories of bulge formation are
primordial or secular evolution scenarios.
While monolithic collapse 
(Eggen et al. \cite{ELS} (ELS)), clumpy collapse
(Kauffmann et al. \cite{kau}, Baugh et al. \cite{bau}), and
inside-out formation (van den Bosch \cite{bosch}, Kepner \cite{kep}) are
proposed for primordial bulge formation, merging processes
(Wyse et al. \cite{wys} and references therein) and dynamical
evolution triggered by bars (e.g. Combes et al. \cite{com90}) are
the basis of secular evolution scenarios.  
However, combinations of both scenarios are also possible (Combes \cite{com99}). 

In the last 20 years several working hypothesis about the origin of b/p bulges
have been developed.
External cylindrically symmetric torques can act on an initially non-rotating
triaxial spheroidal stable system  so that it becomes
box-shaped (May et al. \cite{may}). However, such torques
are ``problematical to relate to some astrophysical counterparts'' 
(Combes et al. \cite{com90}).
Mergers of two disk galaxies are proposed by Binney  \& Petrou (\cite{bin})
and Rowley (\cite{row}) as a possible formation process of b/p bulges. However,
such a scenario requires very special conditions such as the
precise alignment of the spin and orbital angular momenta of
the two galaxies (Bureau \cite{bur}).
A more common process to form b/p bulges could be
the accretion of satellite galaxies, 
although an oblique impact angle of the satellite is needed
(Binney  \& Petrou \cite{bin}, Whitmore \& Bell \cite{whi}) and
the mass of the accreted satellite is constrained.
If it is too large, the stellar disk would be disrupted
(Barnes \cite{bar}, Hernquist \cite{her}). 
The origin of b/p bulges by bars was first proposed by
Combes \& Sanders (\cite{com81}). This
theory was strongly supported ten years later by the studies
of Combes et al. (\cite{com90}), Raha et al. (\cite{rah}), and
Pfenniger \& Friedli (\cite{pfe})
dealing with N-body simulation of disk galaxies.

The connection between b/p bulges and the presence of a bar is difficult
to prove by surface photometry, because b/p bulges are observable only in
almost edge-on galaxies. However, there is photometric
evidence in a few edge-on galaxies from
cuts parallel to the major axis (de Carvalho \& da Costa \cite{deca},
Dettmar \& Ferrara \cite{df}) and 
in two intermediately inclined galaxies (Bettoni \& Galletta
\cite{bet}, Quillen et al. \cite{qui}) pointing to bars. Additionally,
kinematic observations of around 20 b/p bulges with
characteristic bar signatures  in velocity fields and in orbit analyses
(e.g. Kuijken \& Merrifield \cite{kui})
are reported. 
In addition, our statistics (Paper I)
have shown that the overall frequency of barred galaxies
($\sim$ 55\,\%, Paper I,
Knapen et al. \cite{kna2000}) 
is able to explain the high fraction of b/p bulges.

On the other side observational evidence 
for accreted material in b/p bulges is
found by Bureau \& Freeman (\cite{bf}) for a few b/p bulges.
Indirect evidences are substructures and asymmetries in some
special b/p bulges
(Dettmar \cite{det89}, Dettmar \& Barteldrees \cite{det90a},
L\"utticke et al. \cite{lue2000b}, hereafter Paper III).
Combes et al. (\cite{com90}) and Shaw (\cite{sha93}) conclude from
their investigation of b/p bulges, listed by Jarvis (\cite{jar}),
de Souza \& dos Anjos (\cite{sa}), and Shaw (\cite{sha93}),
that they show less evidence for significant merging
and are likely not the result of recent accretion events.
However, b/p bulges can indirectly originate from interaction and finally 
accretion by triggering the formation of bars
(Noguchi \cite{nog}, Gerin et al. \cite{ger}, Walker et al. \cite{wal}),
which then actually build up the b/p structures (Fisher et al. \cite{fis},
Mihos et al. \cite{mih}).
Therefore the search for bars is one of the best ways
to discriminate between the proposed evolutionary scenarios of b/p bulges. 

The essential advantage of investigations in the NIR is
the small extinction  by dust at these wavelengths
(Rieke \& Lebowsky \cite{rie}, Knapen et al. \cite{kna91}).
Therefore the influence of the dust lane in the 
plane of edge-on galaxies is largely reduced by NIR observations,
especially in the $K$-band.
In this way the structure of a galaxy near the plane in radial
as well as in vertical direction can be analysed
and thereby bars detected (Dettmar \& Ferrara \cite{df}).
In addition it is even possible to determine parameters of bars,
e.g. their thickness and  projected length. 
Therefore it is possible to investigate
the connection between bars and shapes of bulges.


\section{Observation and data reduction}

Observations were carried out at the
1.5m TIRGO/Gor\-ner\-grat in 1994/95 (run identification: TI) 
with NICMOS3 ARNICA (0.92$\arcsec$/pixel),
the 1.2m telescope at Calar Alto
in June 1996 (CA4) and
 June 1997 (CA2) with MAGIC (Rockwell NICMOS3 array)
using the high resolution mode (1.2$\arcsec$/pixel),
the 2.2m telescope at Calar Alto during two runs
in January 1997 (CA3) and March 1998 (CA1) also equipped 
with MAGIC (hr mode: 0.64$\arcsec$/pixel), and
at ESO/La Silla with the 2.2m telescope in May 1998 (E1)
with IRAC-2b (NICMOS3 array)
using the LC objective (0.49$\arcsec$/pixel). To the resulting sample of
56 galaxies four galaxies kindly provided by Emsellem
(E2: ESO/La Silla 2.2m telescope, April 1998, IRAC-2b, LC objective)
are added.
The sample was selected from the list of edge-on galaxies
with bulge classification (Paper I) 
according to the allocated observing time, matching the field size, and
covering all bulge types.
Filters and typical on-source integration times for each observing run are
listed in Table \ref{obs} and the  final sample
is given in Table \ref{gal}. 
At ESO/La Silla and TIRGO the $K'$- resp. $K$-band filters were used.
Due to large thermal emission in $K$ for MAGIC we observed
at Calar Alto mainly in $H$.
Using different filters we can show 
that the bulge shape in the NIR
is independent from the observed wavelengths (Paper I).

\begin{table}[hbtp]
\caption{Observational data of the NIR}
\label{obs}
\begin{center}
\begin{tabular}{l|cccc}
(1) & (2) & (3) & (4) & (5)  \\
obs. run & f & int. time & seeing & num. of gal. \\
\hline \hline
TI & $K$ & 40 &  1.8$\arcsec$ & 10 \\
\hline
CA1 & $J$ & 20 & 1.5$\arcsec$ & 2 \\
CA1 & $H$ & 32 & 1.5$\arcsec$-2.5$\arcsec$ & 20 \\
CA1 & $K$ & 25 & 1.5$\arcsec$ & 2 \\
\hline
CA2& $H$ & 22 & 2$\arcsec$-3$\arcsec$ & 18\\
\hline
CA3 & $J$ & 5 & 1.8$\arcsec$ &4\\
CA3 & $H$ & 5 & 1.8$\arcsec$ &4\\
CA3 & $K$ & 6 & 1.8$\arcsec$ &4\\
\hline
CA4 & $H$ & 30 & 2.8$\arcsec$ & 4 \\
\hline
E1 & $K'$ & 20 & 1.5$\arcsec$ & 5 \\
\hline
E2 & $K'$ & 12 & 1.5$\arcsec$ & 4 \\
\end{tabular}
\end{center}

Notes: \\
Col. (1): Observing run; abbreviation see text.\\
Col. (2): Filter.\\
Col. (3): Typical integration time on source in minutes. \\
Col. (5): Number of observed galaxies; some galaxies were observed twice.

\end{table}

Data reduction is performed with standard techniques for NIR images:
dark subtraction, sky background subtraction, bad pixel correction,
and flatfielding (CA1-4: sky flats, E1+2: dome flats)
(see e.g. MAGIC Observer's Guide \footnote{http://www.caha.es/CAHA/Instruments} 
or ESO-MIDAS User Guide, Volume B
\footnote{http://www.eso.org/projects/esomidas}).
The sky background frames
 are determined by offsets,
after each object frame,
integrating sky images with the same time  as used for the object images.
Combining the individually reduced object frames leads to
the final image of the galaxy. For the photometric calibration of the images
infrared standard stars (Elias et al. \cite{eli}; Persson et al. \cite{per}) 
are used
giving a photometric error of $\Delta\!\sim\!0.3$ mag. 
The data obtained at TIRGO
are calibrated using published integrated
aperture data. 

\section{Analysis of the data}

Following Wakamatsu \& Hamabe (\cite{wak}) ``bumps'', which are symmetric 
to the minor axis,
in cuts  along or parallel to the major axis
of edge-on disk galaxies
can be associated with bars. This is indicated by the fact
that the bump becomes more and more insignificant at larger distances from
the plane (D'Onofrio et al. \cite{dono}).
Using this technique bars in edge-on galaxies
NGC 4762
(Wakamatsu \& Hamabe \cite{wak}), 
NGC 1381 (de Carvalho \& da Costa \cite{deca}),
NGC 5170 (Dettmar \& Barteldrees \cite{det90}), NGC 4302 
(Dettmar \& Ferrara \cite{df}), NGC 3250B, NGC 5047, NGC 6771, IC 4767 
(L\"utticke \cite{lue99}),
and at a marginal level in
NGC 128 (D'Onofrio et al. \cite{dono}) have been found.
However,
for edge-on galaxies these radial cuts are obviously
hampered by dust  near the plane. Therefore previous studies
using optical observations are limited
to early type galaxies with low dust content 
or the resulting evidence
for a bar remain uncertain (NGC 5170). The influence of dust
is significantly reduced in the NIR enabling the identification of bars
also in late type spirals
(Dettmar \& Ferrara \cite{df}).

\begin{figure}
\centering
\resizebox{\hsize}{!}{\psfig{figure=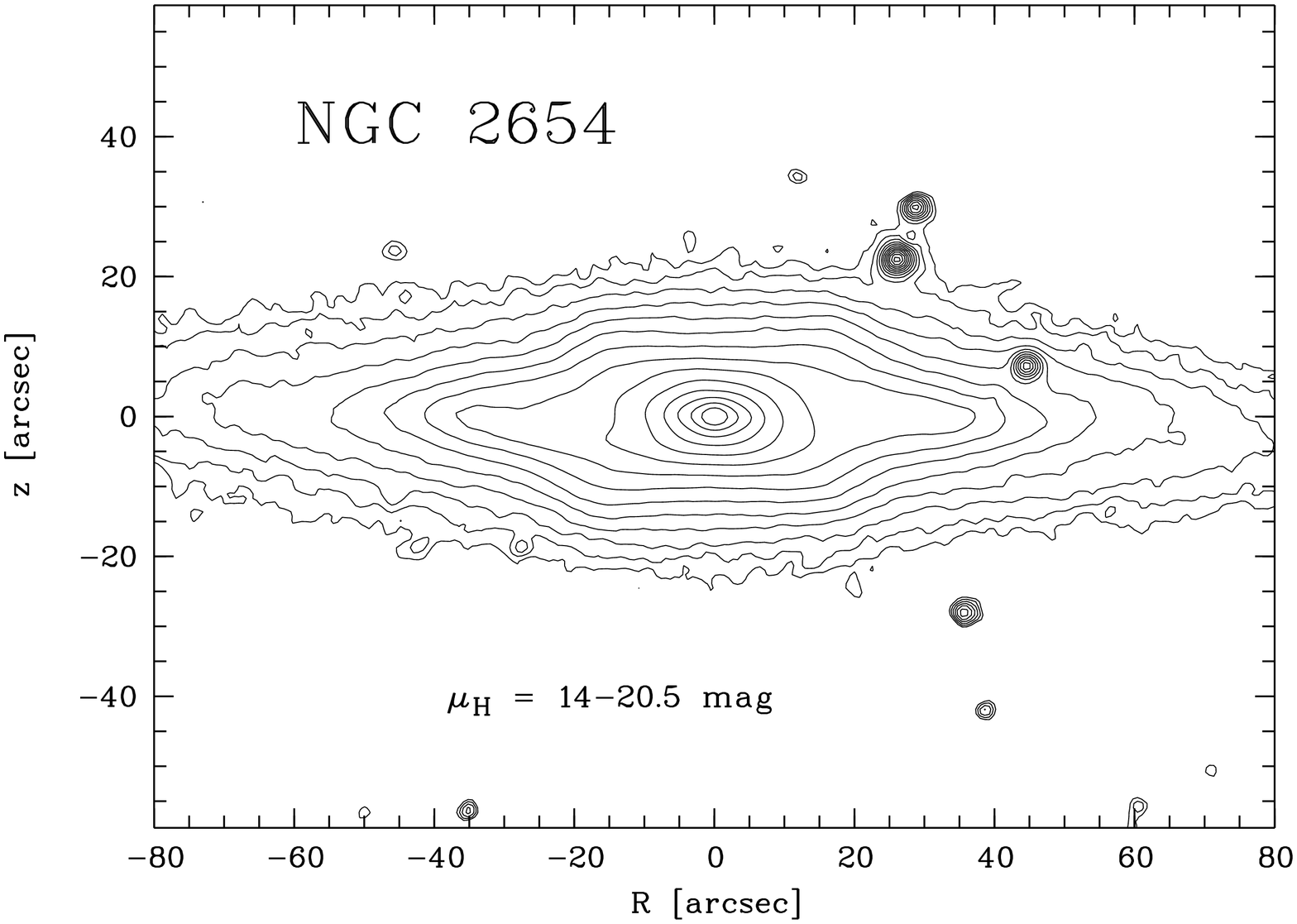,width=8.4cm,angle=270}}
\resizebox{\hsize}{!}{\psfig{figure=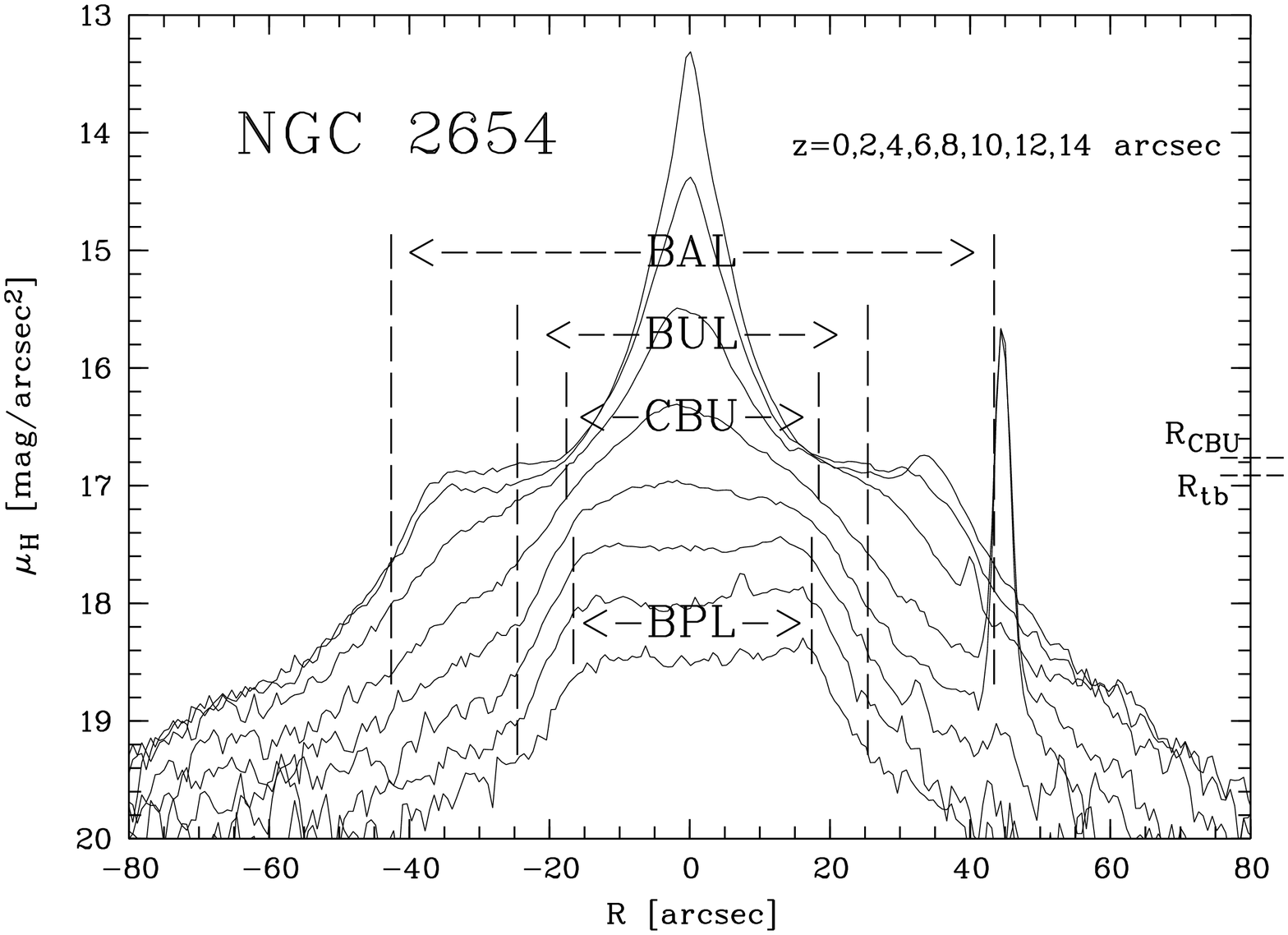,width=8.4cm,angle=270}}
\caption{Top: The peanut bulge of NGC 2654 in the NIR. CA1, 25min in $H$.
Bottom: In cuts along and parallel to the major axis of NGC 2654 the parameter
BAL, BAT, CBU, BUL, and BPL can be determined as explained in the text.
The bar signature can be detected in cuts of the range z = [$-3\arcsec, +3\arcsec$].
The flat bump has a gradient
$\nabla_{\rm bump}\!=\!(16.90 - 16.75)$  mag $* (34 - 18)^{-1} {\rm arcsec}^{-1}\!=\!0.01$ mag arcsec$^{-1}$
(strong bar signature).
The feature at the top of the bump at R = +34$\arcsec$  can be associated with a
spiral arm.
}
\label{N2654}
\end{figure}

\begin{figure}
\centering
\psfig{figure=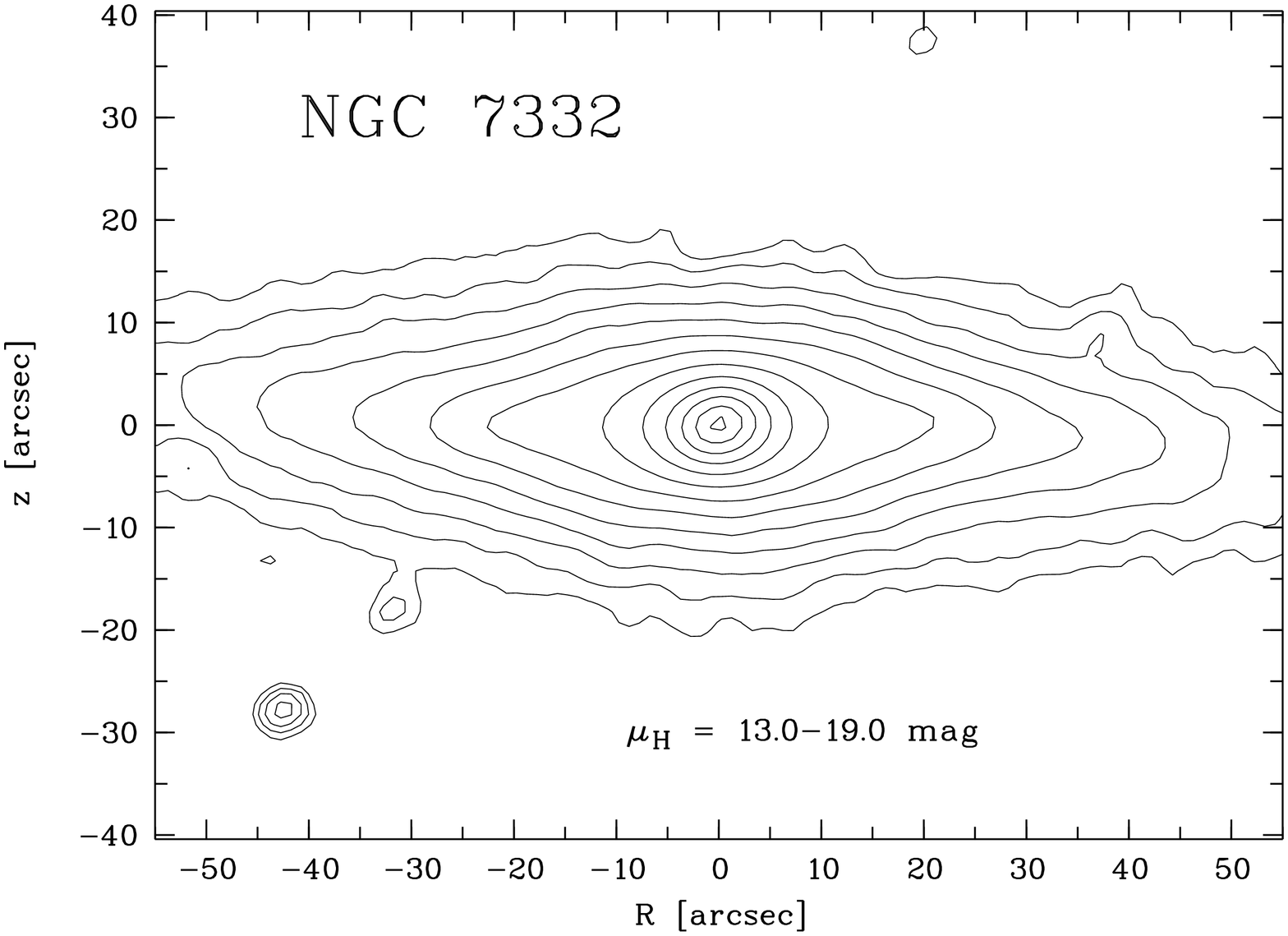,width=8.4cm,angle=270}
\psfig{figure=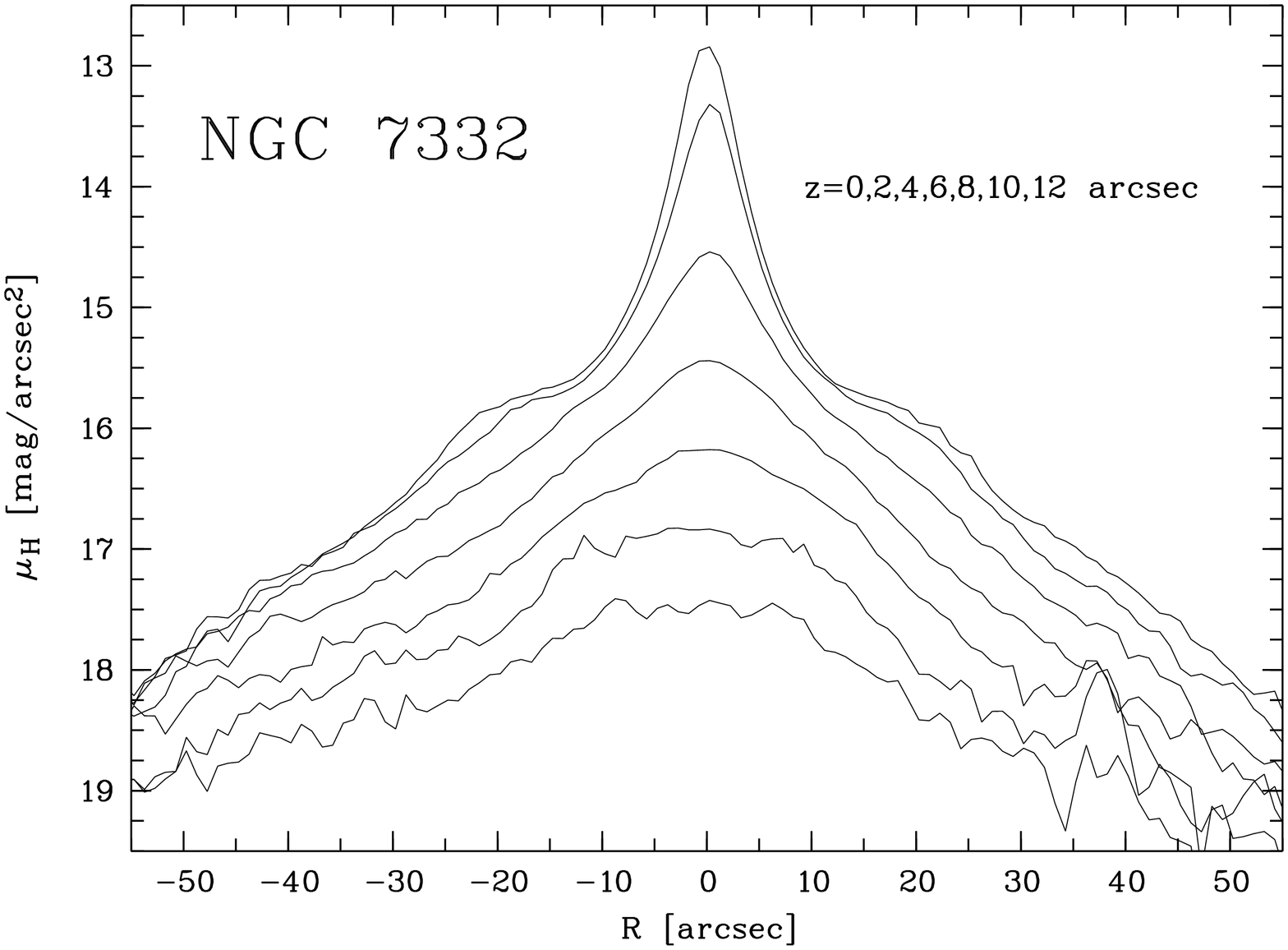,width=8.4cm,angle=270}
\caption{Top: Contour plot of NGC 7332. CA4, 20min in $H$.
Bottom: Cuts along and parallel to the major axis reveal a decreasing bump:
$\nabla_{\rm bump} \sim 0.03$ mag arcsec$^{-1}$ (weak bar signature).}
\label{N7332}
\end{figure}

\begin{figure}

\centering
\psfig{figure=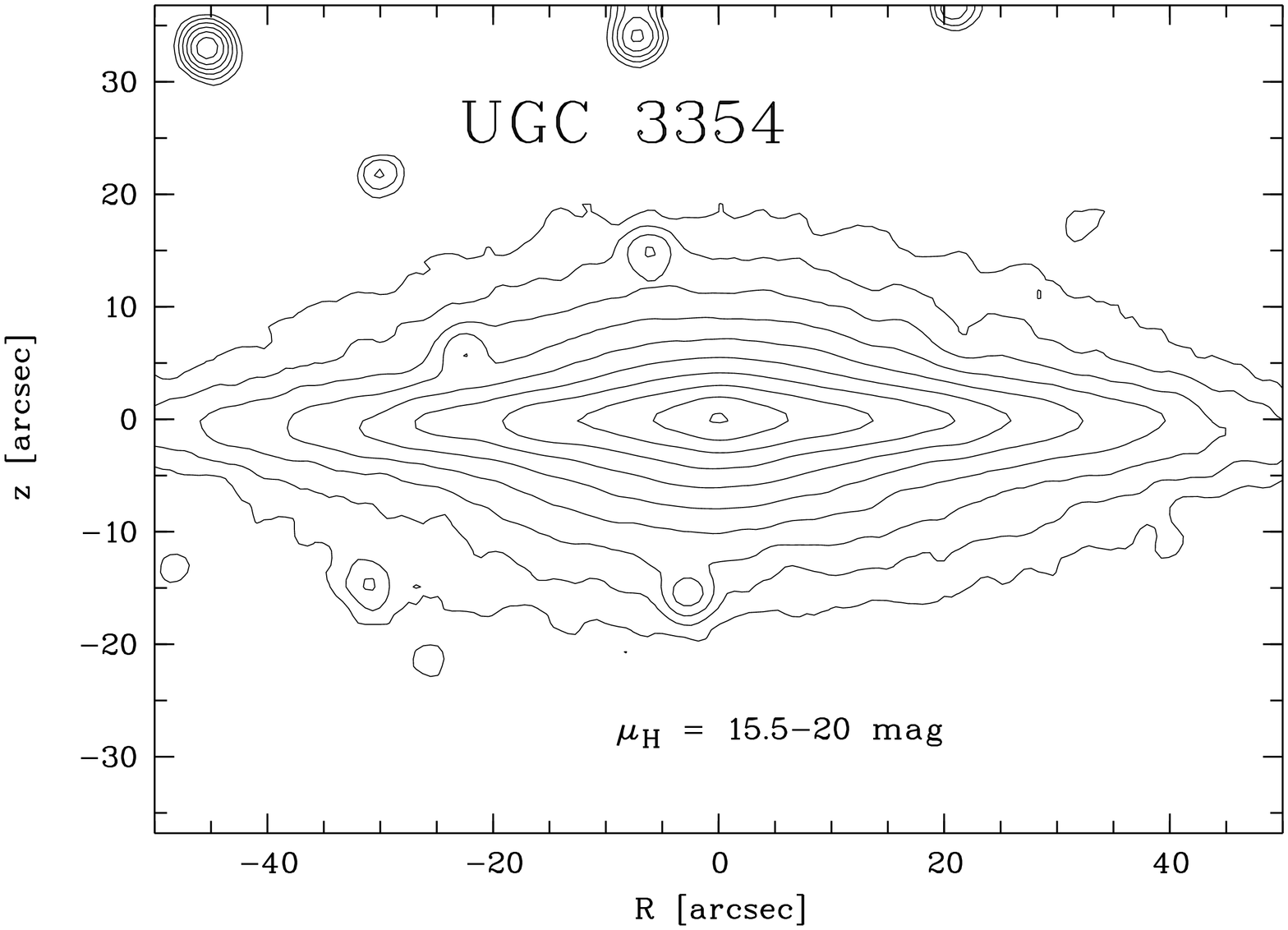,width=8.4cm,angle=270}
\psfig{figure=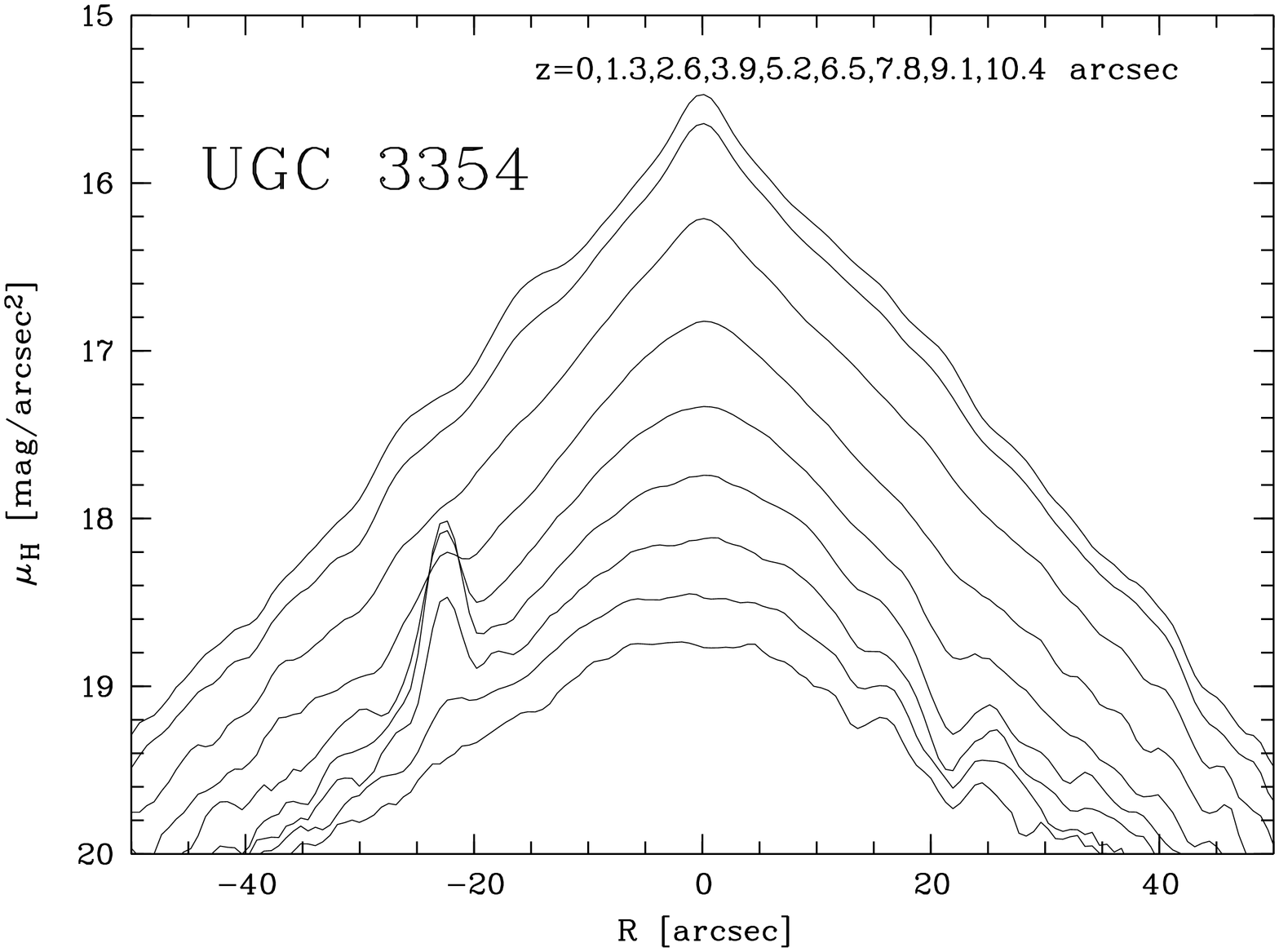,width=8.4cm,angle=270}
\caption{Top: Contour plot of UGC 3354. CA1, 35min in $H$.
Bottom: There is no bar signature visible
in the cuts along and parallel to the major axis.}
\label{U3354}
\end{figure}

Therefore we have searched for bumps, which are signatures of a bar,
in radial cuts of the galaxies  in our NIR-sample.
Additionally, the shape of the bumps in the cuts along and parallel to
the major axis can be divided in profiles with a flat part
and in profiles decreasing from the center of the galaxy throughout.
For these purposes we have determined 
the surface brightness at the radial position
of the top of the bump
($R_{\rm tb}$) and  at the end of the
central bulge ($R_{\rm CBU}$).
We call the feature in the cuts ``strong bar signature'', if
the gradient
\begin{displaymath}
 \nabla_{\rm bump} = \left | \frac{\mu (R_{\rm tb}) - 
  \mu (R_{\rm CBU})}{R_{\rm tb} - R_{\rm CBU}} \right |
\end{displaymath}
is smaller
than 0.015 mag arcsec$^{-1}$ (Fig. \ref{N2654} and \ref{N1175}).
Otherwise the bump is called ``weak bar signature'' (Fig. \ref{N7332}).
The error of this gradient is $\pm$0.003 mag arcsec$^{-1}$.
An example for a galaxy without a bar signature is given in Figure \ref{U3354}.

For galaxies showing a bar signature
several characteristic quantitative parameters are derived from the
cuts parallel to the major axis (for illustration see Fig. \ref{N2654}):
\begin{itemize}
\item {\bf BAL}:
The projected bar length is marked in the cut along the major axis
by an increasing light distribution towards the center compared to the
radial exponential light distribution (Pohlen et al. \cite{poh}) of the
disk.
\item {\bf BAT}: 
The bar thickness is defined by the vertical extent of cuts 
showing the bar signature.
\item {\bf CBU}:
We introduced the length of the central bulge as the length of the 
structure being brighter than the bar signature.
\item {\bf BUL}:
The bulge length is marked 
by the increasing light distribution over the exponential disk well
above the bar.
\item {\bf BPL}: The length of the b/p structure is
distance between the maxima of the b/p distortion. This parameter is obviously
only definable for peanut bulges.
\end{itemize}
However, taking into account seeing effects 
the measured values are only upper limits.

For 17 galaxies of our sample it is not possible to search for
bar signatures. 
Sometimes the heavy dust extinction near the plane,
even in the NIR, prevents a final analysis.
In other galaxies
extra features (e.g. stars or possible spiral arms), 
asymmetrical structures, or  low signal-to-noise ratios would lead to a
too uncertain verification of a bar.
Therefore the resulting sample is reduced to 43 galaxies.

\section{Results}

\begin{table*}[hbtp]
\caption{Results of the structure analysis}
\label{gal}
\begin{center}
\begin{tabular}{l|lcccccccccc}
\multicolumn{1}{c|}{(1)} & (2) & (3) & (4) & (5) & (6) & (7) & (8) & (9) & (10) & (11) & (12)   \\
\multicolumn{1}{c|}{Object} & bulge &
 bar  & BAL & BAT & BUL & CBU & BPL & BAL/ & BAL/ &  BAL/ & CBU/  \\
& type &&[$\arcsec$]&[$\arcsec$] &[$\arcsec$]&[$\arcsec$]&[$\arcsec$]& 
BAT & BUL & BPL & BPL \\
\hline 
IC 2233 & 3 & ? & & & & &  \\
NGC 2683 & 2 & ? & & & & &  \\
NGC 3628 & 1 & ? & & & & &  \\
NGC 4026 & 4 & WB & 64 & 6 & 62 & 26 & &10.7 & 1.0&\\
NGC 4217 & 4 & -- & & & & &  \\
NGC 4302 & 2 & SB & 76 & 8 & 46 & 22 & & 9.5 & 1.7&\\
NGC 4634 & 4 & ? & & & & &  \\
NGC 4762 & 4 & WB & 120 & 8 & 104 & 30 && 15 & 1.2 &  \\
NGC 5908 & 4 & -- & & & &   \\
NGC 5775 & 3 & ? & & & &  \\
\hline
UGC 3354 &4 & -- & & & & & \\
UGC 3458 & 3 & -- & & & & & \\
NGC 2654 & 1 & SB & 86 & 6 & 50 & 36 & 34 &  14.3 & 1.7 & 2.5 & 1.1\\
NGC 2862 & 1 & SB & 74 & 5 & 42 & 24 & 26 & 14.8 & 1.8 & 2.8 & 0.9\\
NGC 2939 & 3 & WB & 34 & 4 & 26 & 12 & & 8.5 & 1.3&\\
UGC 5904 & 3 & ? & & & & & \\
NGC 3669 & 2 & ?  & & & & & \\
NGC 3986 & 1 & ?  & & & & & \\
NGC 3987 & 1 & ?  & & & & & \\
IC 755 & 3 & ?  & & & & & \\
NGC 4469 & 1 & SB & ? & 24 & 96 & 60 & 60 &   & &  & 1.0\\
NGC 5014 & 4 & ? & & & & & \\
NGC 5166 & 1 & SB & 44 & 3 & 26 & 16 & 14 & 14.7 & 1.7 & 3.1 & 1.1\\
UGC 8737 & 3 & ? & & & & & \\
NGC 5470 & 4 & -- & & & & &  \\
NGC 5492 & 4 & -- & & & & & \\
UGC 9759 & 2 & -- & & & & & \\
UGC 9841 & 2 & WB & 46 & 4 & 28 & 18 & & 11.5 & 1.6&\\
UGC 10205 & 3 & -- & & & & & \\
NGC 6310 & 2 & SB &62 & 7 & 34 & 24 & & 8.9 & 1.8&\\
\hline
NGC 5859 & 2 & SB & 52 & 8 & 34 & 20 & & 6.5 & 1.5&\\
NGC 5864 & 3 & WB & 66 & 10 & 50 & 36 & & 6.6 & 1.3&\\
UGC 9853& 4 & -- & & & & & \\
UGC 9858 & 4 & WB & 30 & 4 & 26 & 10 & & 7.5& 1.2&\\
NGC 5981 & 4 & -- & & & & & \\
UGC 10227 & 3 & WB & 34 & 4 & 20 & 14 & &8.5 & 1.7&\\
UGC 10447& 4 & -- & & & & & \\
NGC 6361  & 4 & ? & & & & &  \\
UGC 11455 & 4& -- & & & & & \\
UGC 11893 & 4 & -- & & & & & \\
UGC 11973 & 2 & ? & & & & & \\
NGC 7264& 4 & -- & & & & & \\
NGC 7339 & 3 & ? & & & & & \\
UGC 12506& 4 & -- & & & & &  \\
NGC 7640& 2 & ? & & & & &  \\
\hline
NGC 1175 & 2+ & SB & 54 & 5 & 36 & 20 & & 10.8 & 1.4 &\\
NGC 2424 & 1$^*$ & SB & 76 & 6 & 36 & 28 & 32 & 12.7 & 2.1& 2.4 & 0.9\\
NGC 3098 & 3 & WB & 30 & 3 & 26 & 18 & &10 & 1.2 &\\
NGC 3762 & 3$^*$ & -- & & & & & \\
\hline
NGC 5707 & 4 & -- & & & & &  \\
NGC 7332 &  3 & WB & 56 & 6 & 40 & 24 & & 9.3 & 1.4 &\\
\end{tabular}
\end{center}
\end{table*}

\begin{table*}[hbtp]
Table \ref{gal} continued.
\begin{center}
\begin{tabular}{l|lcccccccccc}
\multicolumn{1}{c|}{(1)} & (2) & (3) & (4) & (5) & (6) & (7) & (8) & (9) & (10) & (11) &
(12)   \\
\multicolumn{1}{c|}{Object} & bulge &
 bar  & BAL & BAT & BUL & CBU & BPL & BAL/ & BAL/ &  BAL/ & CBU/  \\
& type &&[$\arcsec$]&[$\arcsec$]&[$\arcsec$]&[$\arcsec$]&[$\arcsec$]& 
BAT & BUL & BPL & BPL \\
\hline

ESO 560- 13 & 2+ & ? & & & & &  \\
NGC 2788A & 1 & SB & 90 & 6 & 38 & 30 & 28 & 15  &2.4 & 3.2 & 1.1\\
ESO 506- 3 &  3$^*$ & -- & & & & &  \\
ESO 443- 42 & 1 & SB & ? & 5& 46 & 24 &  24 &  & &  & 1.0  \\
IC 4745 & 3$^*$ & -- & & & & & &&&  \\
\hline
IC 760 & 3 & WB & 36 & 3 & 26 & 16 & &12 & 1.4 &\\
NGC 2310 & 2 & SB & ? & 5 & 44 & 30 & &&&\\
NGC 3203 & 3 & SB & 50 & 4 & 34 & 24 & & 12.5 & 1.5 &  \\
NGC 6771 & 1 & SB & 70 & 5 & 44 & 28 & 32 & 14 & 1.6 & 2.2 & 0.9\\

\end{tabular}
\end{center}

Notes: \\
Col.(1): Galaxies are ordered by RA for each observing run separated
by lines (TI, CA1, CA2, CA3,
CA4, E1, and E2 ).
If a galaxy is observed twice, only the data with higher quality
are listed.\\
Col.(2): Bulge type as defined in Paper I
in the NIR, except $^*$: 
optical bulge type (Paper I), because signal-to-noise ratio in the NIR image
is too low. \\
Col.(3): Bar signatures: -- = no signature, WB = weak bar signature,
SB = strong bar signature, ? = uncertain (see text)
\\
Col.(4) - (7):
Abbreviations see text. ? = BAL is larger than
the size of the image.
\\
Errors of the measurements : BAL and BUL $\sim$\,20\,\%,
BAT $\sim$\,30\,\%, CBU and BPL $\sim$\,10\,\%. \\
\end{table*}

The derived bar signatures and parameters for our sample are
presented in Table \ref{gal}.
Furthermore, we have computed characteristic ratios of the parameters
(Table \ref{gal}).

\begin{figure}[htbp]
\centering
\resizebox{\hsize}{!}{\psfig{figure=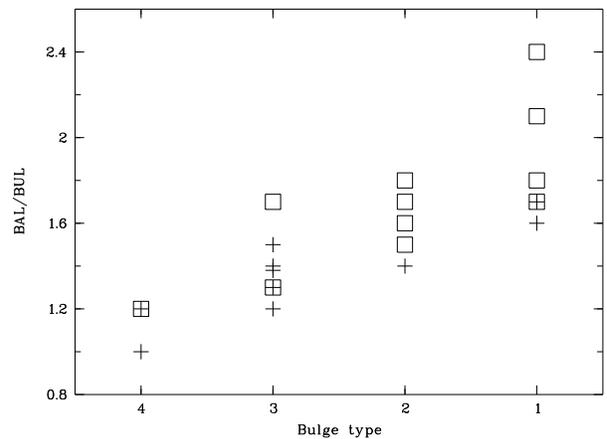,width=8.4cm,angle=270}}
\caption{The bulge type (in the NIR) seems to be correlated with the ratio of
the projected length of the bar (BAL) and the bulge length (BUL).
Crosses: S0-Sab; squares: Sb-Scd.}
\label{balbul}
\end{figure}

In Paper I
we have visually
classified bulge shapes in three 
types of b/p bulges
({\bf 1}: peanut-shaped, {\bf 2}: box-shaped, {\bf 3}: nearly box-shaped)
and one elliptical type ({\bf 4}).
Regarding the ratio of the projected bar length to the bulge length
(BAL/BUL) we measure a clear trend with the bulge type (Fig. \ref{balbul}):
The more prominent the b/p type, the larger  the ratio.
The mean value for type 1 bulges is BAL/BUL $\!=\!1.9 \pm 0.3$,
type 2: $1.6 \pm 0.2$, type 3: $1.4 \pm 0.2$, and type 4: $1.1 \pm 0.1$.
However, we are dealing with small number statistics,
${\rm N}\!=\!21$, and the scatter 
is relatively high. This could be explained
by the different morphological types
contributing to these statistics, because early type galaxies 
have smaller ratios than late types (Fig. \ref{balbul}).
Additionally, according to Martin (\cite{mar}),
Elmegreen \& Elmegreen (\cite{2elm95}), and Elmegreen et al. (\cite{elm})
the relative bar length
($R_{\rm bar}/R_{25}$) has also a high scatter pointing to a large
range of bar lengths.

Two other characteristic ratios, CBU/BPL and BAL/BPL, are only related 
to peanut bulges.
They can be compared with
predictions of N-body simulations
and thereby test the evolution scenario of b/p bulges proposed in
these simulations (see next section). 
The mean value of
BAL/BPL is $2.7 \pm 0.3$ (Table \ref{gal}).
A surprising result for CBU/BPL is
the mean value of 1.0 with the small scatter of 0.1 
(Table \ref{gal} and Fig. \ref{N2654}). 

\begin{table}[ht]
\caption{Connection between b/p bulges and bar signatures}
\label{connec}
\begin{center}
\begin{tabular}{c|ccc}
classified & \multicolumn{3}{c}{number of galaxies with} \\
bulge type & no sig. & WB $^1$ & SB $^2$\\
\hline 
1 & 0 & 0 & 8 \\
2 & 0 (1) & 2 & 5 \\
3 & 1 (4) & 5 & 1 \\
4 & 13 & 3 & 0 \\
\end{tabular}
\end{center}
Notes:\\
galaxies in brackets are excluded (see text) \\
$^1$: Weak bar signature\\
$^2$: Strong bar signature\\

\end{table}

Five out of the six galaxies
without a bar signature (Table \ref{connec}),
but with a b/p bulge are thick boxy bulges (TBBs) 
(Dettmar \& L\"utticke \cite{dl}).
This group of galaxies shows
remarkable morphological features such as large asymmetries and
substructures 
and the box-shape is more prominent in the outer part of the bulge,
possibly pointing to a different formation scenario.
Therefore this group of b/p bulges will be studied separately in a
forthcoming paper (Paper III).
The remaining galaxy (NGC\,3762) without  bar signature and  b/p bulge
is classified as a weak type
3 bulge on an optical image (Paper I).
Excluding the five TBBs from
Table \ref{connec} there is an
almost one-to-one correspondence  between b/p 
bulges and bar
signatures. The fraction of b/p bulges with such a signature is 95\,\%
whereas over 80\,\%  of the non-b/p bulges show no bar signature. Additionally,
it should be mentioned that the three galaxies (NGC\,4026, NGC\,4762, and
UGC\,9858) with a bar signature and
without a b/p bulge 
have the smallest
BAL/BUL ratios in our sample (Table \ref{gal} and Fig. \ref{balbul}).
Table \ref{connec} reveals a significant correlation on one side between  
strong bar signatures  and
prominent b/p bulges (type 1 + 2) and on the other side
between  weak bar signatures
and less pronounced b/p bulges of type 3.
This strong dependence is shown on a 1.0\,\% significance-level by 
a $\chi ^2$-test. 
We do not find a correlation between the Hubble type
and the weak or strong  bar signature.

\begin{figure}[htbp]

\resizebox{\hsize}{!}{\psfig{figure=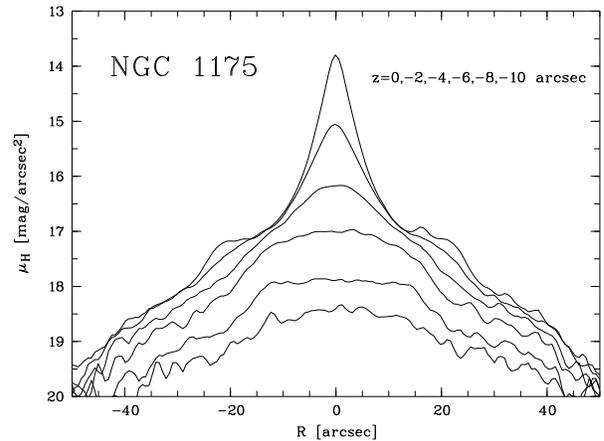,width=8.4cm,angle=270}}
\caption{Cuts along and parallel to the major axis of NGC 1175
 show  strong bar signatures. This points
to the existence of a bar, which is 
not listed in the RC3. CA3, 6 min in $H$.}
\label{N1175}
\end{figure}

For NGC\,1175 (Fig. \ref{N1175}) and NGC\,3203 we measure 
a strong bar signature.
Therefore the classification as unbarred systems within the RC3 can be
replaced. Since the bar classification for edge-on galaxies is
difficult and lead to misclassifications in
galaxy catalogues based on photographic plates, 
our method of bar detection points at a solution for
this problem.  

\section{N-body simulation}
\subsection{Model}
The origin of b/p bulges based on the evolution of bars is shown
in several N-body simulations (e.g. Combes et al. \cite{com90},
Pfenniger \& Friedli \cite{pfe}).
To check the consistency of simulations 
with the results of our observations,
we have investigated a representative simulation of a barred galaxy.
For this purpose a particle position file of the
final stage of a
N-body simulation of a barred galaxy in three dimensions
was kindly made available by Daniel Pfenniger.
The simulation uses a particle-mesh code with 200.000 particles
described in
Pfenniger \& Friedli (\cite{pfe}). 
They choose axisymmetric initial conditions which
are likely to develop
a bar instability on their own and an initial mass distribution
corresponding to a superposition of two axisymmetric
Miyamoto-Nagai disks. 
The provided binary file represents the final time step of
their simulation at T = 5000
($\sim$\,5 Gyrs) in which a bar is formed.
The length unit is chosen to be one kpc (Pfenniger \& Friedli \cite{pfe}).
For the transformation of 
the three-dimensional mass distribution of the particles
to a two-dimensional surface brightness distribution
we assume a constant M/L ratio and
take into account the inclination ($i$)
of the galaxy and the aspect angle ($\phi$) of the bar. An edge-on bar
is denoted with $\phi\!=\!90^{\circ}$, 
while $\phi\!=\!0^{\circ}$ refers to an end-on bar.

\begin{figure*}
\centering
\psfig{figure=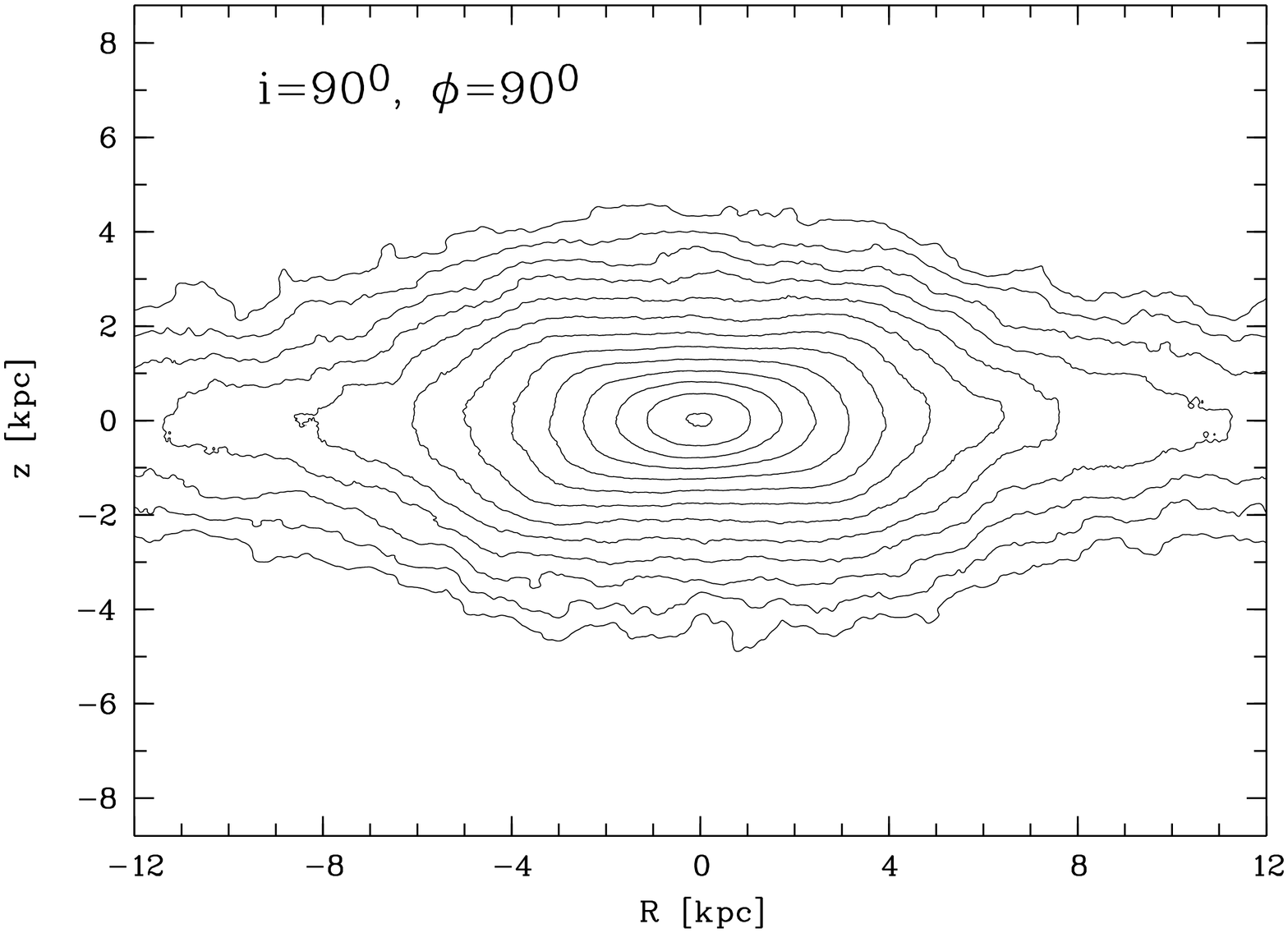,width=7.8cm,angle=270}
\psfig{figure=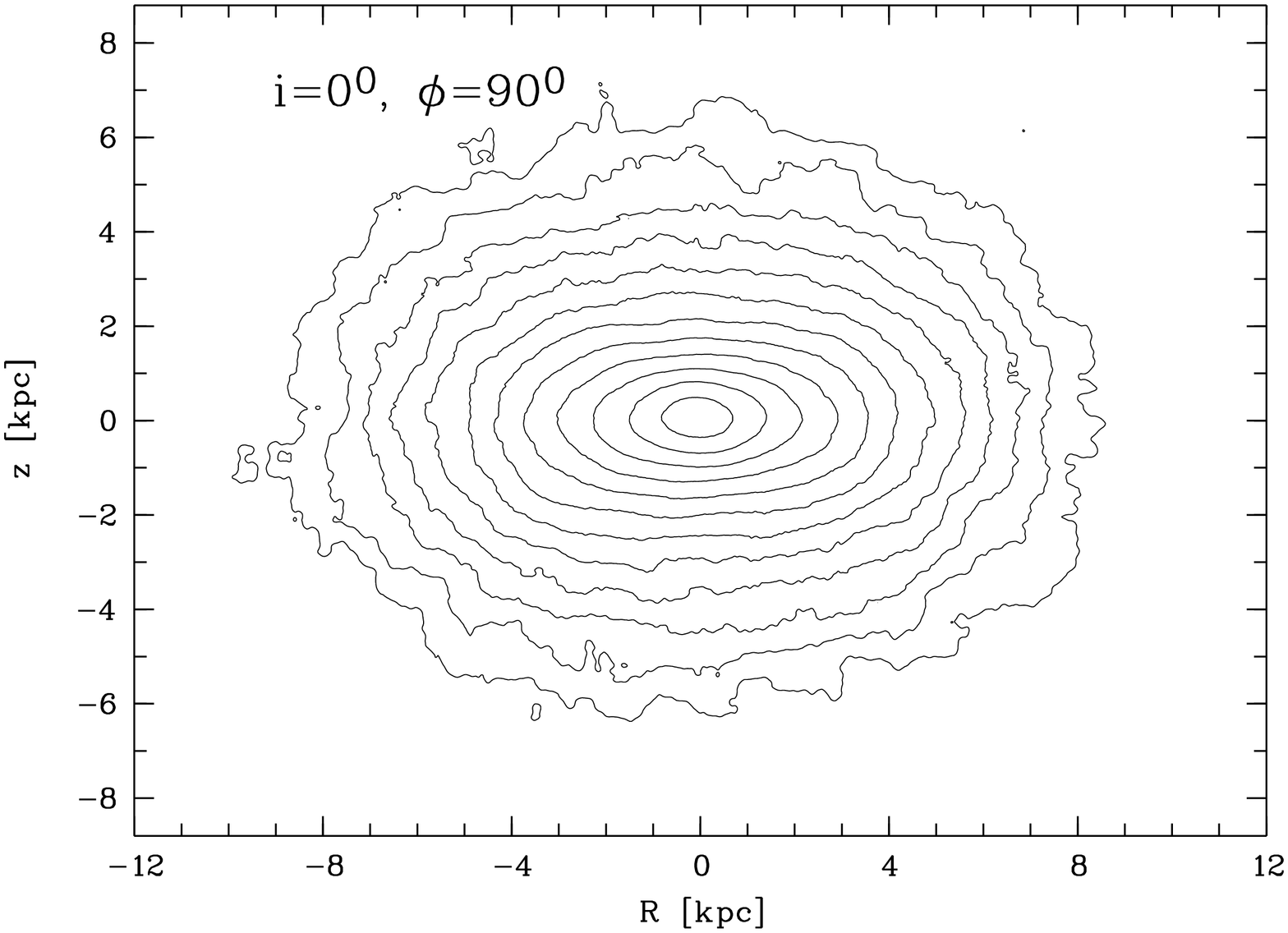,width=7.8cm,angle=270}
\psfig{figure=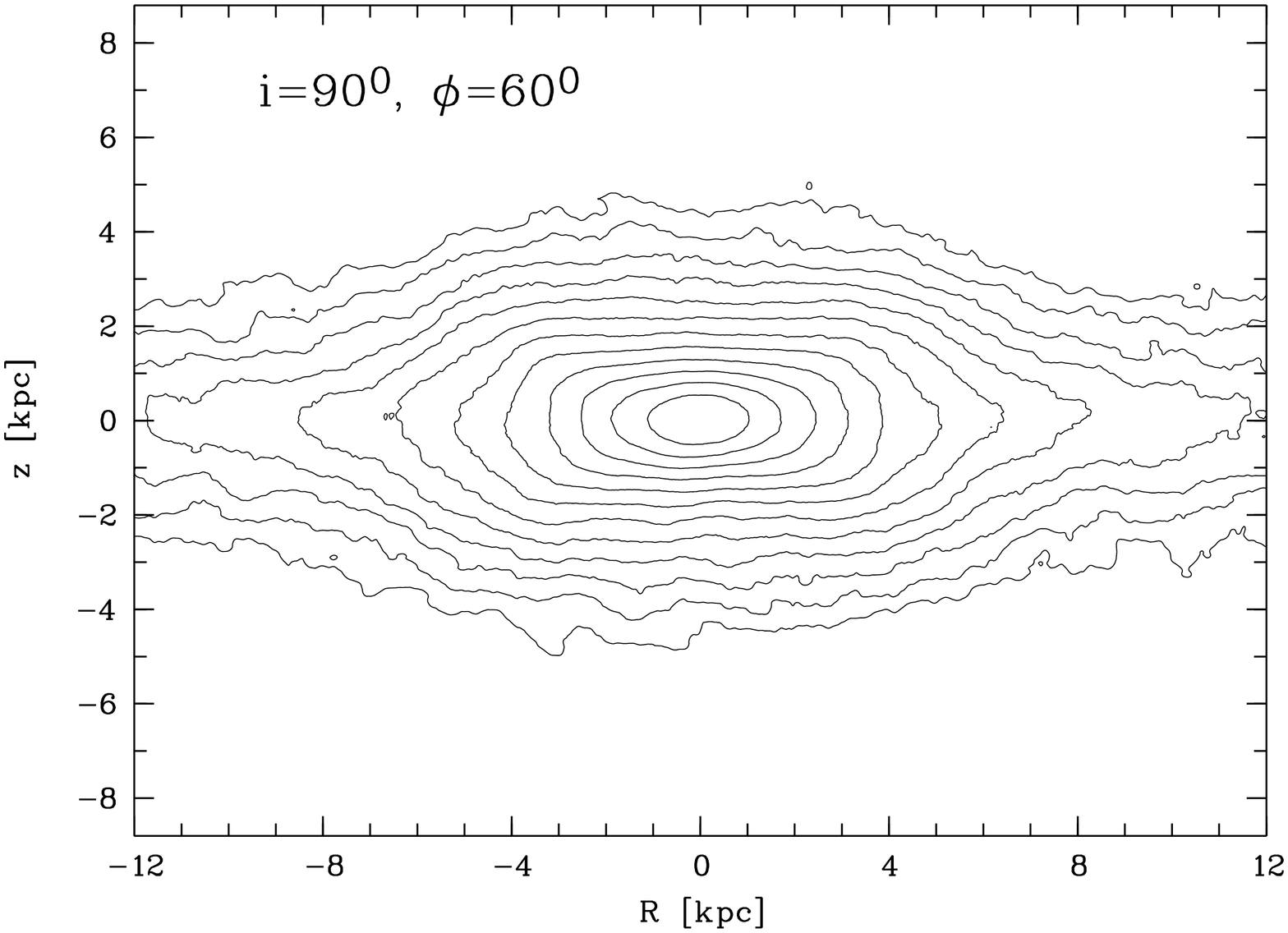,width=7.8cm,angle=270}
\psfig{figure=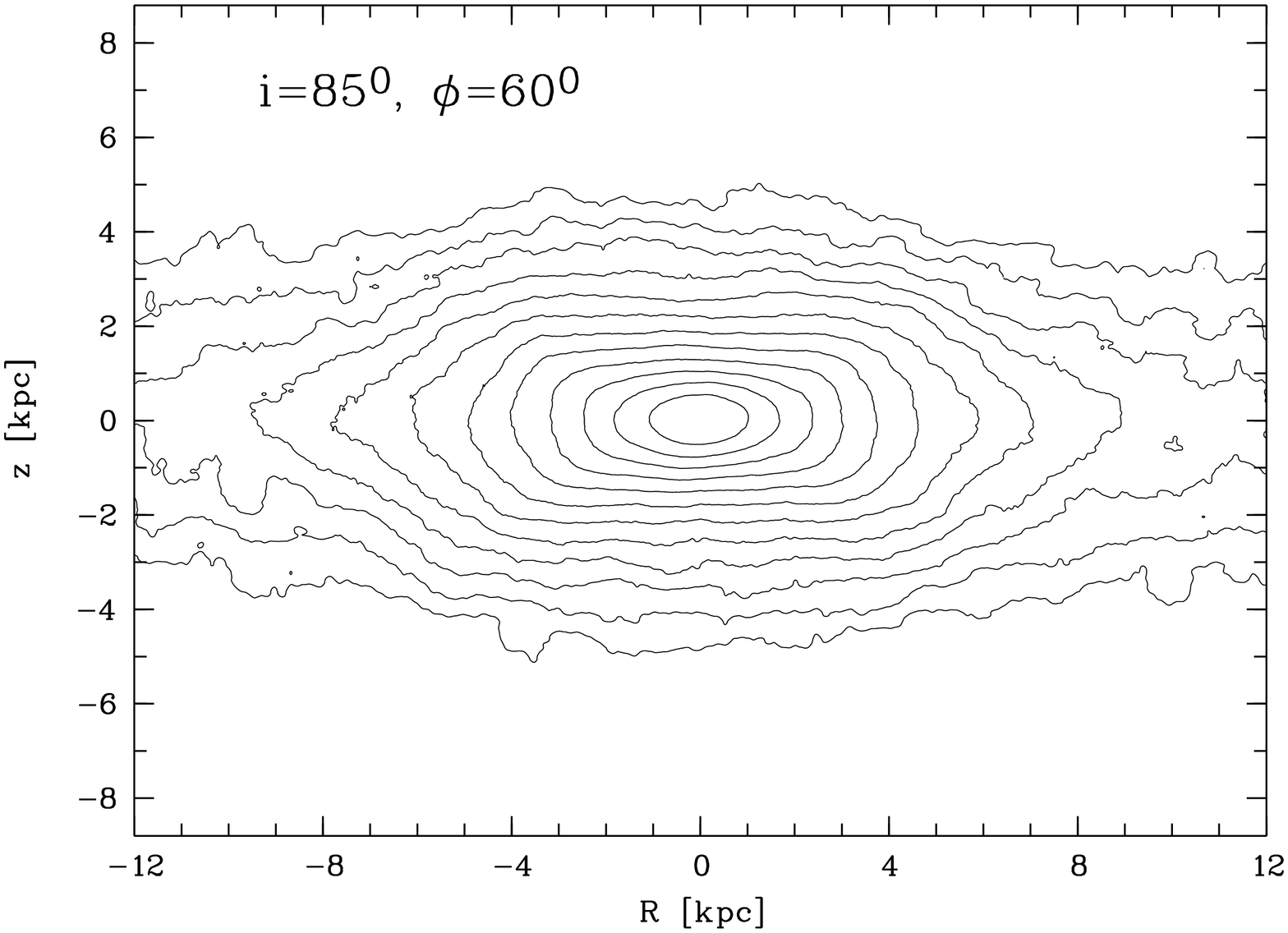,width=7.8cm,angle=270}
\psfig{figure=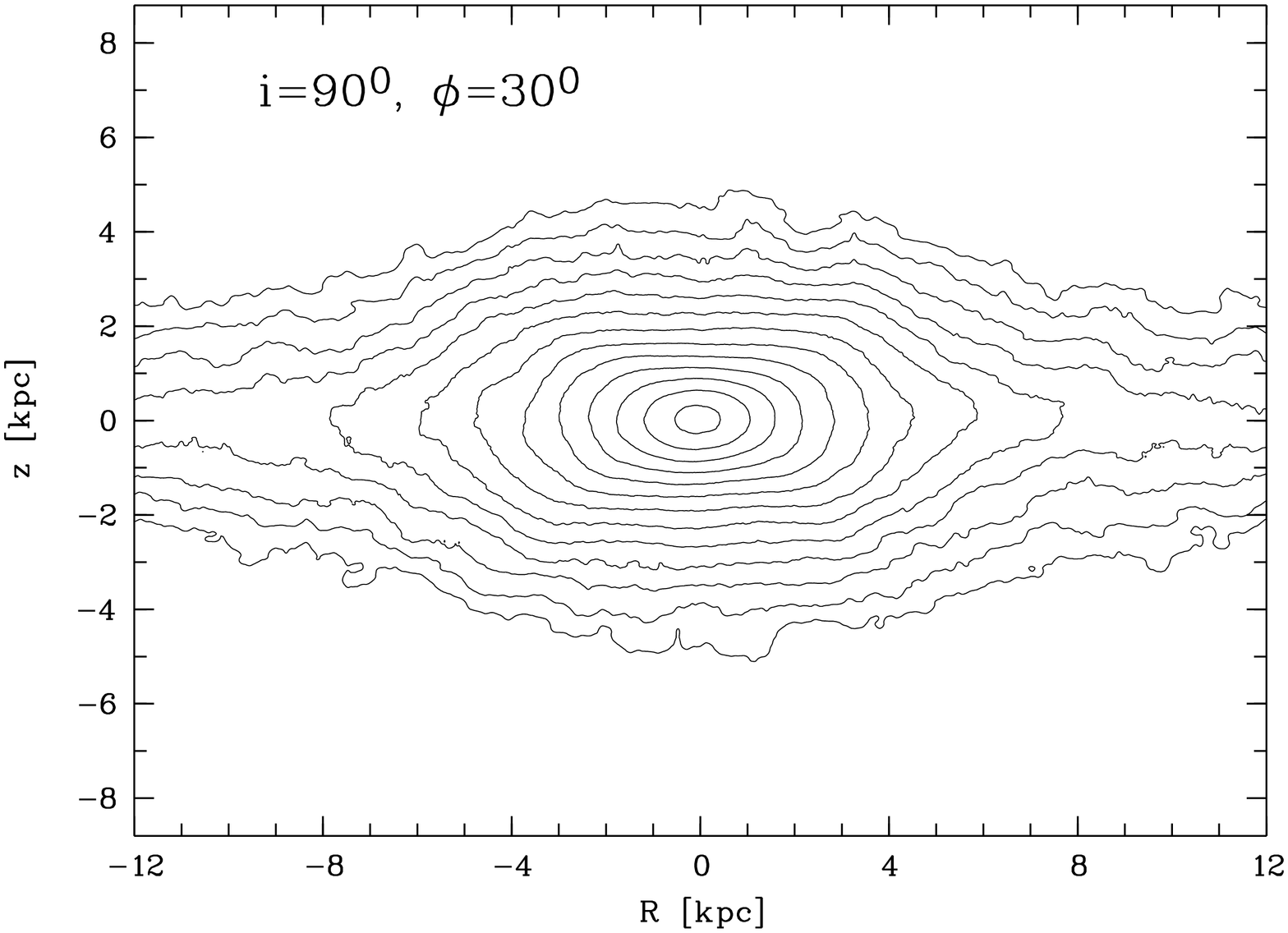,width=7.8cm,angle=270}
\psfig{figure=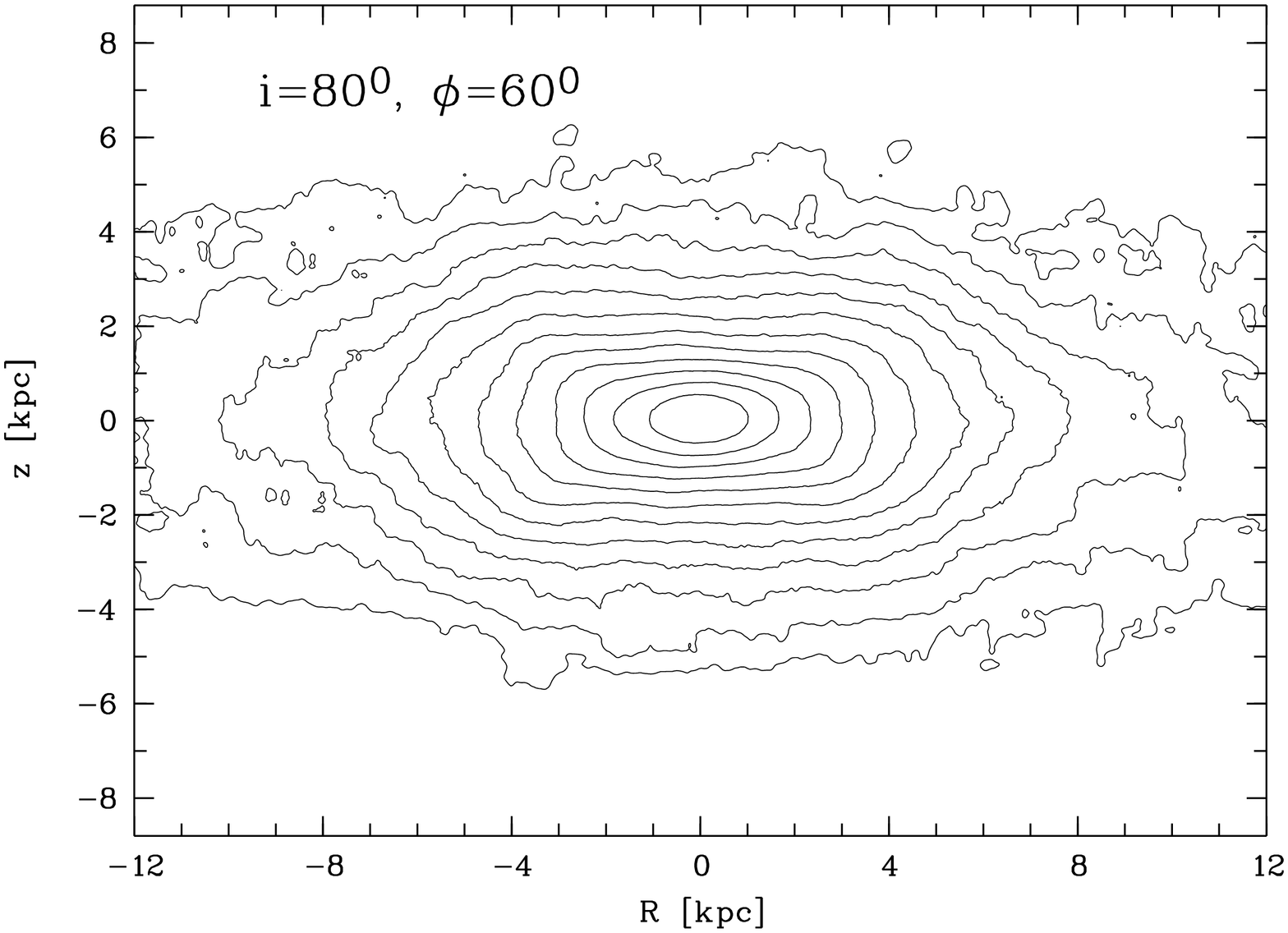,width=7.8cm,angle=270}
\psfig{figure=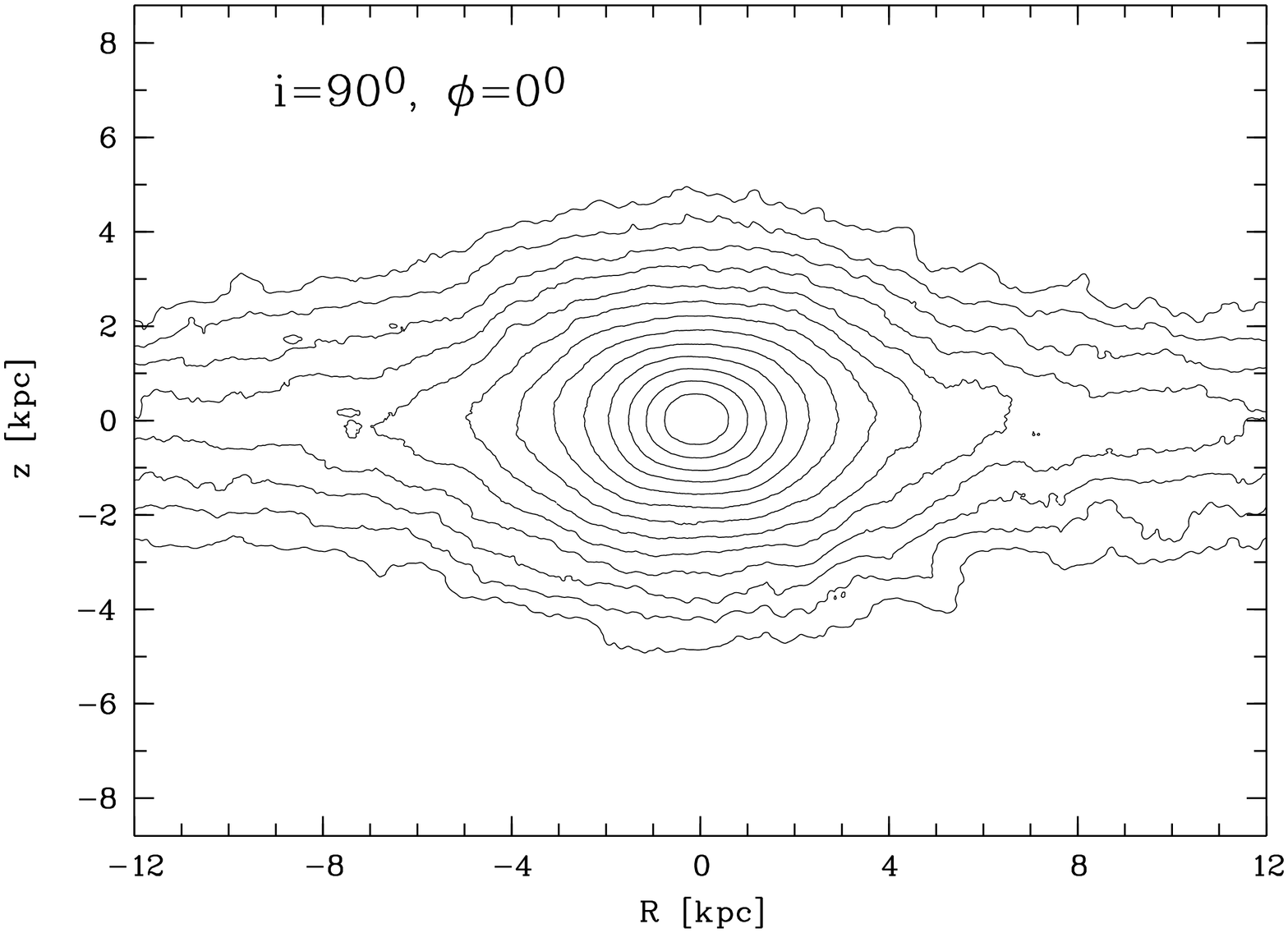,width=7.8cm,angle=270}
\psfig{figure=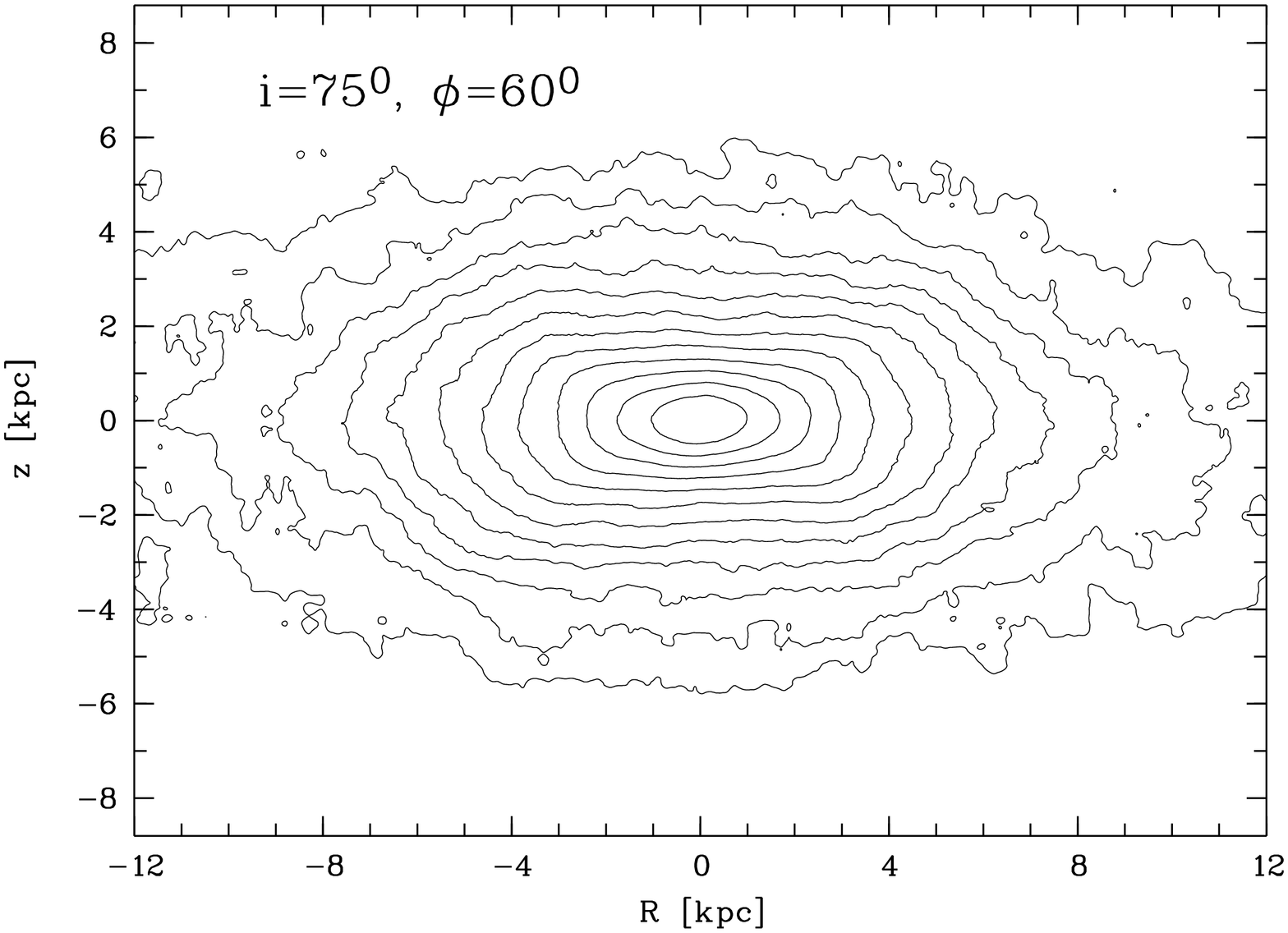,width=7.8cm,angle=270}

\caption{Contour plots of the projected density depending on the galaxy
inclination ($i$) and the aspect angle of the bar ($\phi$).
Left side: The galaxy is edge-on,
while the aspect angle of the bar decreases. Right side: Top:
Galaxy is face-on. Bottom: The aspect angle
of the bar is constant, while the inclination of the galaxy decreases.
Same density isophotes are used in all contour plots.
}
\label{simu}
\end{figure*}

\subsection{Classification of the bulges}

In the investigated N-body simulation the peanut shape of bulges 
can be detected down
to a galaxy inclination of $\sim$\,$75^{\circ}$, if the bar is edge-on.
If the galaxy is exactly edge-on, peanut bulges can be seen 
down to an
aspect angle of the bar of $\sim$\,$77^{\circ}$. 
In general the simulation shows that
the degree of boxiness of the bulge, reflected by the bulge type,
 is weakened with decreasing
inclination of the galaxy (Fig. \ref{simu}, Table \ref{sim}).
At inclinations lower than $\sim$\,75$^{\circ}$ it is not possible
to distinguish between structures perpendicular to the galactic plane
and the bar itself. This value was also previously found by
Shaw et al. (\cite{sdb}) as limit for the observation of b/p bulges.
Regarding galaxy inclinations above this limit the b/p morphology 
is detected in the simulation 
whenever $\phi$ is larger than $\sim$\,20$^{\circ}$.
In this way $\sim$\,77\,\% of projection angles of bars result in 
b/p bulges
(Table \ref{sim}).
Combes et al. (\cite{com90}) find in their simulations a similar 
percentage (85\,\%) for ``favourable'' projection angles to see b/p bulges
in galaxies with an inclination between 80$^{\circ}$ and 90$^{\circ}$.

Since the frequencies of the simulated bulge types (Table \ref{sim}, Col. 4)
are derived  only from
a barred galaxy, these frequencies must be modified to compare them
with our statistics of bulge types (Paper I) derived from all edge-on galaxies,
with bar and without.
Taking this into account the percentages for the bulge types of the 
simulated galaxy have to be 
multiplied with the observed fraction of barred galaxies.
Adding afterwards 
the fraction of unbarred galaxies 
to the percentage of bulge type 4  --- 
N-body simulations (Pfenniger \& Friedli \cite{pfe})
suggest that unbarred galaxies have such a bulge type ---
leads to the modified frequencies of the bulge types 
(Table \ref{sim}, Col. 5 and 6).
The errors of these frequencies depend on
the uncertainties of the limits in the aspect angle of the bar ($\phi$)
 (up to $\pm 4^{\circ}$) and have values of
$\sim$\,2.5\,\% for type 1 and 4 bulges and
$\sim$\,5\,\% for type 2 and 3 bulges. 
The observed frequencies of the bulge types 
(Paper I; type 1: 4.1\,\%,
type 2: 15.7\,\%, type 3: 25.2\,\%, and type 4: 55.0\,\%)
are very similar 
(inside the errors)
to the modified frequencies computed with a 
fraction of 55\,\% barred galaxies. This percentage is revealed
by studies from optical surveys (e.g. RC3)
(Ho et al. \cite{ho}, Knapen et al. \cite{kna2000}, Paper I).
However, the higher percentage of 70\,\% derived from NIR studies 
(Mulchaey \& Regan \cite{mul}, Knapen et al. \cite{kna2000},
Eskridge et al. \cite{esk}) seems to be
more reliable, since the NIR-light traces the dominant, old stellar
population and is relatively unaffected by dust.
Using this fraction for computing the modified frequencies,
there is a difference of $\sim$\,10\,\% between observed and
modified frequencies of galaxies with b/p bulges (type 1-3).
There are less b/p bulges observed as computed from 
the investigated simulation coupled with NIR observations.

The existence of galaxies with non-b/p bulges, but with
bars (Tab. \ref{gal})
are consistent with the simulation where almost end-on orientated bars
lead to elliptical bulges.

\begin{table}[hbtp]
\caption{Bulge types in the N-body simulation}
\label{sim}
\begin{center}
\begin{tabular}{c|c|c|c|c|c}
(1) & (2) & (3) & (4) & (5) & (6) \\
bulge & $i$ & $\phi$ & fre- & mod.$_{55\,\%}$ & mod.$_{70\,\%}$ \\
type & & & quency & frequency & frequency \\
\hline
1 & 90$^{\circ}$ & 77$^{\circ}$-90$^{\circ}$ & 14.4 \% & 7.9 \%  & 10.1 \% \\
2 & 90$^{\circ}$ & 47$^{\circ}$-76$^{\circ}$ & 33.3 \% & 18.3 \% & 23.3 \% \\
3 & 90$^{\circ}$ & 16$^{\circ}$-46$^{\circ}$ & 34.5 \% & 19.0 \% & 24.1 \% \\
4 & 90$^{\circ}$ & 0$^{\circ}$-15$^{\circ}$ & 17.8 \% & 54.8 \%  & 42.5 \% \\
\hline
1 & 85$^{\circ}$ & 81$^{\circ}$-90$^{\circ}$ & 10.0 \% & 5.5 \% & 7.0 \% \\
2 & 85$^{\circ}$ & 52$^{\circ}$-80$^{\circ}$ & 32.2 \% & 17.7 \% & 22.5 \% \\
3 & 85$^{\circ}$ & 18$^{\circ}$-51$^{\circ}$ & 37.8 \% & 20.8 \% & 26.5 \% \\
4 & 85$^{\circ}$ & 0$^{\circ}$-17$^{\circ}$ & 20.0 \% & 56.0 \% & 44.0 \% \\
\hline
1 & 80$^{\circ}$ & 85$^{\circ}$-90$^{\circ}$ & 5.6 \% & 3.1 \% & 3.9 \% \\
2 & 80$^{\circ}$ & 57$^{\circ}$-84$^{\circ}$ & 31.1 \% & 17.1 \% & 21.8 \% \\
3 & 80$^{\circ}$ & 20$^{\circ}$-56$^{\circ}$ & 41.1 \% & 22.6 \% & 28.8 \% \\
4 & 80$^{\circ}$ & 0$^{\circ}$-19$^{\circ}$ &  22.2 \% & 57.2 \% & 45.5 \% \\
\hline
1 & 75$^{\circ}$ & 89$^{\circ}$-90$^{\circ}$ & 1.1 \% & 0.6 \% & 0.8 \% \\
2 & 75$^{\circ}$ & 66$^{\circ}$-88$^{\circ}$ & 25.6 \% & 14.1 \% & 17.9 \% \\
3 & 75$^{\circ}$ & 30$^{\circ}$-65$^{\circ}$ & 40.0 \% & 22.0 \% & 28.0 \%\\
4 & 75$^{\circ}$ & 0$^{\circ}$-29$^{\circ}$ & 33.3 \% & 63.3 \% & 53.3 \% \\
\end{tabular}
\end{center}

Notes:\\
Col. (1): Bulge type as defined in Paper I.\\
Col. (2): Inclination of the simulated galaxy. \\
Col. (3): Aspect angle of the bar ($90^{\circ}$ = edge-on and 
$0^{\circ}$ = end-on).\\
Col. (4): Frequency of the bulge type of the simulated galaxy, 
viewed under different
aspect angles, at a constant inclination.
 \\
Col. (5): Modified frequency of the bulge type of the  simulated galaxy,
viewed under different aspect angles, at a constant inclination
taking into account
that only 55\,\% of all galaxies are barred
and the remaining galaxies
without a bar (45\,\%) have a bulge of type 4. \\
Col. (6): As Col. (5), but with 70\,\% barred and 30\,\% unbarred galaxies
\\
\end{table}

\subsection{Quantitative parameters}

Along the bar major axis of 
the simulated galaxy, measured in the face-on case ($i\!=\!0^{\circ}$),
the decrease of the surface density is exponential
(Pfenniger \& Friedli \cite{pfe}) and the length of the bar can be determined to
$\sim$\,20 kpc.  This length exactly fits
inside the radius of corotation which occurs at
10 kpc (Pfenniger \& Friedli \cite{pfe}).
The axis ratio of the bar is $b/a\!=\!0.7 \pm 0.1$  
(Fig. \ref{sim}, top right)
and is inside the range of observed values (0.2 - 1)
(Martin \cite{mar}).

\begin{figure}[ht]
\begin{center}
\resizebox{\hsize}{!}{\psfig{figure=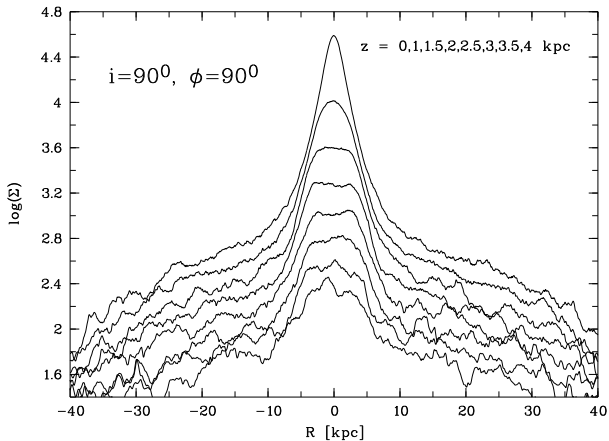,width=8.4cm,angle=270}}
\end{center}

\caption{Cuts along and parallel to the major axis of the simulated galaxy
with $i\!=\!90^{\circ}$ and $\phi\!=\!90^{\circ}$. Cuts with the
same distance from the major axis are averaged.
BPL $\!=\!8$ kpc and BUL $\!=\!16$ kpc can be derived from the cuts, while a bar
signature is not detectable.  BAL $\!=\!20$ kpc is determined at $i\!=\!0^{\circ}$,
see Fig. 2 in Pfenniger \& Friedli (\cite{pfe}).}
\label{simcut}
\end{figure}

Regarding the galaxy edge-on ($i\!=\!90^{\circ}$) with an
edge-on orientated bar ($\phi\!=\!90^{\circ}$) is the best way to derive 
the quantitave bar parameters (introduced in section 3) of the simulated galaxy,
because  bar and
b/p structures are most prominent at this geometry.
However, the cuts along and parallel to the major axis do not reveal the
typical bar signature (Fig. \ref{simcut}) which is visible in the
NIR observations. The three components bulge, bar, and central part
of the bulge cannot be distinguished as it is possible in the
observations (cf. Fig. \ref{N2654}, \ref{N7332}, and \ref{N1175}).
Due to the lack of a bar signature
the bar thickness as well as the central bulge length (CBU)
cannot be determined.
The ratio of BAL/BUL $\!=\!1.25 \pm 0.2$, taking
into account BAL $\!=\!20$ kpc and BUL $\!=\!16$ kpc (Fig. \ref{simcut}), is  
not consistent with the observed ratios for peanut bulges ($1.9 \pm 0.3$) 
(Fig. \ref{balbul}), while
BAL/BPL $\!=\!2.5 \pm 0.2$ (BPL $\sim$\,8 kpc; Fig. \ref{simcut}) is consistent with
observed ratios in peanut bulges (BAL/BPL $\!=\!2.7 \pm 0.3$).
While in observations the positions of the
maxima of the b/p distortion
coincide with the limits of the central bulge,
in the N-body simulation the positions of the maxima are not related
to any other features in the profiles.

\subsection{Relevance of the simulated galaxy for observations} 

The differences between the galaxy derived from the N-body simulation
and the observed galaxies likely result from the initial conditions of
the simulated model. Pfenniger \& Friedli (\cite{pfe}) 
are just using two merged disks with the same rotation properties
instead of an independent bulge component in their simulation.
The inner disk represents the ``bulge'' and is chosen
with the main purpose
``to have sufficient freedom to shape the initial
rotation curve''. 
Therefore the initial resemblance with a real bulge is
only approximate. 
Using such a bulge component in the simulation 
could also be the reason for
the lack of the bar signature when the galaxy is seen edge-on, since
the inner disk could cover the signature.
Another explanations for the fact that no bar signature is visible
could lay in the limited spatial resolution
or in the lack of gas in the simulation, which
influences the bar structure.
It is also possible that the exponential decrease
of the bar surface brightness and the smooth end of the bar prevent
the detection of a bar signature. 

Additionally, the investigated simulated galaxy
is only a single, although representative, example of a barred galaxy.
Variations of the initial conditions such as the bulge-to-disk ratio,
Q parameter, gas fraction and distribution, and halo structure  
end up in different bar
strenghts, shapes, thickness, lenghts, and density profiles in
N-body simulations (e.g. Friedli \& Benz \cite{friben}).
Therefore conclusions drawn from comparisons between our
statistics of observed bulge types (Paper I) and results from 
the simulated galaxy, which is investigated in this study,
are limited.
Future studies have to analyse recently published
 simulations of barred galaxies
(e.g. Pfenniger \cite{pfe2000}) to get more relevant comparisons with
observations.

\section{Discussion} 
\subsection{Bars and b/p bulges} 

The strong correlation of the bar signature 
with b/p bulges in our study shows that b/p 
bulges are well explained by dynamical processes in bar potentials.
The ratio of the bar length to the length of the b/p distortion (BPL)
($2.5 \pm 0.2$)
in the simulated barred galaxy
and the observed ratio of BAL/BPL ($2.7 \pm 0.3$)
in galaxies with peanut bulges is consistent with such a scenario.

The correlation of the bar signature with b/p bulges
is previously
claimed, using the same detection method,
by the investigations of  individual galaxies, namely
NGC 1381 (de Carvalho \& da Costa \cite{deca}),
NGC 5170 (Dettmar \& Barteldrees \cite{det90}), and NGC 4302
(Dettmar \& Ferrara \cite{df}).
Many further
evidences for the connection between bars and b/p bulges
are reported in the literature: Bettoni \& Galletta (\cite{bet})
and Quillen et al. (\cite{qui})
find them in their investigation of the intermediately inclined galaxies
NGC 4442 resp. NGC 7582.
Furthermore, N-body simulations 
(e.g. Combes et al. \cite{com90}, Raha et al. \cite{rah},
Pfenniger \& Friedli \cite{pfe}) show that the  evolution of a b/p bulge is
coupled with the development of a bar.

Independent evidence is obtained by observing the gas kinematics 
in N\,{\small II} and H$\alpha$.
The characteristic ``figure-of-eight'' rotation curve which
is a strong signature for the presence of a bar and
is detected in many b/p bulges: e.g. NGC 5746, NGC 5965
(Kuijken \& Merrifield \cite{kui}), NGC 2683 (Merrifield \cite{mer96}), and
additional ten further galaxies by Bureau \& Freeman (\cite{bf}) and
five by Merrifield \& Kuijken (\cite{mer}).
Control samples of non-b/p bulges show no kinematical signature of a bar
(Bureau \& Freeman \cite{bf}, Merrifield \& Kuijken \cite{mer}). 
In two b/p bulges
(ESO 597- 36 and NGC 3390) without kinematic bar signature
Bureau \& Freeman (\cite{bf}) find indications for accretion
events. However, since
the positioning of the slit was difficult (Bureau \& Freeman \cite{bf}),
dust in the galactic plane could affected their observations.
Therefore bars can not be excluded in these two galaxies.
A merger-induced bar formation with
a resulting b/p bulge (Mihos et al. \cite{mih}) could be a possible scenario
for these galaxies.  
The four detected non-b/p bulges:
NGC\,5907 (Miller \& Rubin \cite{mil}), IC\,5096,  ESO\,240-\,11
(Bureau \& Freeman \cite{bf}), and 
NGC 681 (Merrifield \& Kuijken \cite{mer})
with a bar signature in the gas kinematics
can be explained by our analysis of the N-body simulation with
bars which are nearly end-on (see next subsection).

Signatures of the gas kinematics in barred potentials are also
detected in retrograde orbits of NGC 128 (Emsellem \& Arnault \cite{ems}),
the galaxy with one of the most prominent b/p bulges
(D'Onofrio et al. \cite{dono}). 
Direct kinematic evidence for bar streaming motions in two galaxies
with b/p bulges (NGC 3079 and NGC 4388) is presented by
Veilleux et al. (\cite{vei}).
Furthermore, in H\,{\small I} the presence of a bar is also verified in one
galaxy with a peanut bulge analysing the position velocity diagram 
(IC 2531: Bureau \& Freeman \cite{bur97}).
The gas kinematics in the central
region obtained in CO (${\rm J}\!=\!2-1$ line) in the case of NGC 3628
(Reuter et al. \cite{reu}), NGC 891 (Garcia-Burillo \& Guelin \cite{gar}),
and NGC 4013 (Garcia-Burillo et al. \cite{gar99})
again point to the presence of a bar.

The conclusion (often given in the literature, e.g.
Pfenniger \& Friedli \cite{pfe}),
that b/p bulges are thick bars, cannot be supported.
From the observational point of view bars and bulges, if b/p-shaped or not,
are obviously individual components with different lengths
(e.g. Fig. \ref{N2654}). Stellar orbits in a barred potential build
up b/p structures and the bar itself. Therefore b/p bulges are
the composition of a spheriodal bulge and a b/p structure
representing the thick part of the bar.

The non-detection of bars (Table \ref{connec}) in all observed TBBs
($\sim$\,5\,\% of all b/p bulges belong to this group, Paper III) 
is an indication for
the hypothesis that the boxiness of these bulges is not related to the
the presence of a bar. 
This idea gets more convincing support from the fact that
the extent of the box shape is too large to result
from an usual bar potential (Paper III). 
However, the lack of a
bar signature does not necessarily mean that there is no bar, since
weak bars, which have exponential surface brightness profiles
(Elmegreen \& Elmegreen \cite{2elm89}), could remain undetecable by
the here presented search for a bar signature. 

In this way also
the missing bar signature in the simulated barred galaxy is explainable,
because the surface density distribution 
along the simulated bar is exponential and
additionally, the bar axis ratio of this galaxy points to a weak bar
($b/a > 0.6$, Martinet \& Friedli \cite{marfri}).

\subsection{Explanations for various bulge shapes}
The connection between b/p bulges and bars can be 
understood as interplay between resonant bending fed by vertical diffusion
of orbits and instabilities similar to the fire-hose one (Bureau \cite{bur},
Friedli \cite{fri}).
While there is evidence for this connection 
in the most galaxies with b/p bulges,
the question for
the different shape of the b/p bulges remains to be answered.

A correlation between a strong signature of a bar, marked by constant
surface brightness in radial direction (cf. section 3),  and
prominent b/p bulges is found in our study.
Thereby the strong bar signature can be associated with a
flat bar, which is known to be strong (Elmegreen \& Elmegreen \cite{2elm89}).
Therefore the correlation can be interpreted by the assumption
that strong barred galaxies are responsible for prominent b/p bulges,
while galaxies with a weak bar have bulges with a lower degree of boxiness.
N-body simulations by Friedli (\cite{fri}) support this explanation for
different bulge shapes.
In this way it is also possible to explain partly
the observed correlation of BAL/BUL with the bulge type (cf. section 4)
by larger bars in galaxies with prominent b/p bulges,
because flat bars tend to be larger (Elmegreen \& Elmegreen \cite{2elm95},
Elmegreen et al. \cite{elm}).
Since this correlation holds for early as well as for late types,
it cannot be explained by the fact that the same bar will
produce more pronounced b/p bulges, 
the smaller the pre-existing
spheroidal bulge is. 
However, having this in mind 
it is not surprising that the weak bar
of the here investigated simulated galaxy can produce a peanut bulge.
Furthermore, one could expect that peanut bulges more often occur 
in late type spirals, but 
the observed fraction of peanut bulges in galaxies in the range of
S0-Scd is rather constant
and equal to zero in spirals later than Sc (Paper I).  
This result could be interpreted
by a dependence of the bar strength on the Hubble type.
Such a correlation of bars 
is found by Elmegreen \& Elmegreen (\cite{2elm})
and Elmegreen et al. (\cite{elm}).
However, in contrast to them Seigar \& James (\cite{sei}) do not find
that flat (strong) bars are more common in early Hubble types and
exponential (weak) bars in later types.
This statement is in agreement with our
results for the profiles of bar signature concerning the Hubble type,
but it is uncertain, if their results concerning bar morphology detected in
face-on galaxies can be transformed to bar signatures in edge-on galaxies.

Additionally, for the explanation of bulge shapes 
it has to be taken into account
that gas mass and a central black hole effect the shape of bulges.
A significant fraction of gas mass (Berentzen et al. \cite{ber98})
or a supermassive central black hole can weaken the b/p structure 
(Friedli \cite{fri}).

Explaining bulge shapes by aspect angles of bars 
was introduced  by N-body simulations
(Combes et al. \cite{com90},
Pfenniger \& Friedli \cite{pfe}).
Kinematical bar diagnostics in edge-on spiral galaxies using simulations of
families of periodic orbits and hydrodynamical simulations are also consistent
with observations. They confirm the dependence of the boxiness on the aspect
angle of the bar
(Bureau \& Athanassoula \cite{ba}, Athanassoula \& Bureau \cite{ab}).
An exact determination of aspect angles is not possible, because the
envelope of the signature of the outer bar region in the
position-velocity diagram  is too faint
(Athanassoula \& Bureau \cite{ab}).

The strong correlation of BAL/BUL with the bulge type and the
agreement of the observed ratio BAL/BUL only of the  most prominent b/p bulges, 
$1.9 \pm 0.3$, with
that ratio obtained for face-on galaxies, $2.3 \pm 0.7$
(Athanassoula \& Martinet \cite{ath}, Baumgart \& Peterson \cite{baum},
Martin \cite{mar}), 
shows that the differences in BAL/BUL can be interpreted by different
aspect angles of bars. Therefore our analysis
points also to a dependence of the shape of a bulge on the
aspect angle of the bar.
In this way and 
consistent with our investigation of the simulated barred galaxy
peanut bulges result from bars seen nearly edge-on (large
projected bar length, BAL),
boxy bulges from intermediately inclined bars, and elliptical ones
from end-on bars
(small BAL) or they have no bar at all.
Thereby bars detected in non-b/p bulges can be explained as well as
differences between the fraction of barred galaxies (70\,\% in the NIR) 
and galaxies with b/p bulges (45\,\%), although also other
parameters might play a role 
(weak bar, large gas mass, or dominant bulge component).
The dependence of the bulge shape on the aspect angle of the bar
is also supported in our study by the  modified frequencies
of the bulge types  derived in the simulation.
These frequencies are consistent 
with the observed frequencies of the
bulge types, if one keeps in mind that only a single simulated barred galaxy
is compared with all morphological types of disk galaxies.
Larger bulge components in the simulation would weaken the b/p
structures and thereby reduce the modified frequency of b/p bulges. 
Stronger bars would have the contrary effect on the modified frequency.
Additionally, many galaxies of our investigated 
sample with peanut bulge and  strong bar signature
show features at the ends of the bumps, which 
can be associated with spiral arms extending outward
from the ends of a bar (Baggett \& MacKenty \cite{bag}).
Therefore it can be  suggested
that bars of these galaxies are indeed
nearly edge-on.

There exists only a few rough estimates of aspect angles of bars 
in edge-on galaxies from observations.
Wakamatsu \& Hamabe (\cite{wak}) adopt for 
NGC 4762 (bulge type 3
in the optical, type 4 in the NIR; Paper I)
an aspect angle of the bar of $\phi\!=\!33^{\circ} \pm 10^{\circ}$ (although
a lower limit of 21$^{\circ}$ is also possible from their argumentation)
and Garcia-Burillo \& Guelin (\cite{gar})
for NGC 891 (bulge type 2) 
$\phi \sim 45^{\circ}$.
These results are in agreement 
with a dependence of the bulge shape
on the bar aspect angle.


\subsection{A new correlation}
The coincidence of the positions of
the maxima of the b/p distortion with the limits of the central bulge, which
is found here for the first time, cannot finally be explained 
(${\rm CBU/BPL}\!=\!1.0 \pm 0.1$).
Selection effects due to different aspect angles of the bar, which would
influence the length of the b/p distortion (BPL),
can be neglected, since we have shown that peanut bulges, for which
BPL is defined only, can be associated with nearly edge-on bars.
The coincidence points
also to a possible connection between the bar length 
and the length of the central bulge (CBU),
since BPL is connected with BAL.

Therefore possible explanations could be related to secondary bars 
(e.g. Friedli \cite{fri}, Greusard et al. \cite{gre}). 
This idea is supported by the result from N-body simulations
(Friedli \& Martinet \cite{frimar}) that the corotation of the inner bar
roughly coincides with the location of the
horizontal inner Lindblad resonance (ILR).
At this position also the maximum of the elevation of the 
peanut structure occurs, because
the vertical ILR is very close to the horizontal ILR and
the conjunction
of these two resonances build up the peanut structure
(Combes et al. \cite{com90}).
However, the observed ratio of
the bar lengths of the primary to the secondary, which ends near
corotation (Greusard et al. \cite{gre}),
varies between 4.0 and 13.4 (Friedli et al. \cite{fri96}),
respectively 2.7 and 7.1 (Greusard et al. \cite{gre}), while the ratio of
bar length to length of the central bulge
is much smaller and varies only between 2.4 and 3.1.

The central bulge could also be associated with a primordial bulge,
which influences 
the bar structure and the position of the 
horizontal ILR.
Since at this location also 
the peanut structures is most prominent, 
the extent of the b/p distortion would depend on the
mass and length of the primordial bulge.
Such a hypothesis is supported by N-body simulations in which the size
of the evolving bar is directly related to the initial size of the bulge
(Pfenniger \cite{pfe2000}).
In this way the nearly constant ratio of 
bar length or BPL 
to CBU would be a natural result. 

The correlation between CBU and BPL and especially the CBU/BPL
ratio of 1 could also be explained
by  nuclear rings.
They  correspond to the horizontal ILR of bars 
(Buta \& Crocker \cite{but}, 
Buta \& Combes \cite{butcom}) and therefore occur at the same radial position
as the maximum of the elevation of the peanut structure. A nuclear ring
could be responsible for the
increase of the surface brightness,
which we observe in radial cuts
near the plane (Fig. \ref{N2654}, \ref{N7332}, \ref{N1175}),
or the ring is only the border for the central bulge. 
However, we have not seen any structure inside the central bulge
pointing to rings, but it is likely that our resolution is too low
to detect them.

\subsection{Bar thickness} 
Since we know now with high probability the fact
that bars in peanut bulges are edge-on orientated,
it is for the first time possible
to determine the thickness of bars by observations.
From the mean value of BAL/BAT for the galaxies with peanut bulges a ratio
of bar length to thickness of $14 \pm 4$
(taking the error in BAL and BAT into account) is derived.
This value is at the upper limit of the values
estimated in the literature (ranging from 4 to 15,
Sellwood \& Wilkinson \cite{sel}). It must be mentioned that bars
are more extended in the NIR than in the optical
(Friedli et al. \cite{fri96},
Chapelon et al. \cite{cha}). However, from a sample of nine barred galaxies
(Wozniak et al. \cite{woz}, Friedli et al. \cite{fri96}) investigated in
the $I$- and $K$-band we have computed that this effect is only of the order
of $\sim$\,5\%.
Therefore our rather large value for the ratio is not an effect of
the observed wavelengths.

\section{Conclusion}
We have analysed a sample of 
edge-on disk galaxies in the NIR and images derived
from a
N-body simulation of a barred galaxy using cuts parallel to the major axis
and our classification of bulge shapes introduced in Paper I.
Our main conclusions are:

\begin{itemize}

\item Cuts along and parallel to the major axis
of edge-on galaxies show a strong
correlation between bars and b/p bulges. This correlation is for the first time
verified in a larger sample of galaxies (N $\!=\!60$).

\item The investigation of the bar signatures in the NIR reveals 
a correlation between prominent b/p bulges and strong bar signatures
pointing to a dependence of the boxiness of a galaxy on the bar strength.

\item The ratio of the projected bar length (BAL) to the length of
the bulge (BUL) correlates
with the bulge types, i.e. with the bulge shape. 
This correlation is the first evidence from surface photometry
for an
explanation of the different degree of boxiness by the aspect angle of a bar.
The maximum of this ratio for galaxies with
peanut bulges is
consistent with ratios observed in face-on galaxies. 

\item The ratio of BAL to the length of the b/p
distortion (BPL) has a mean value for the observed peanut bulges of
$2.7 \pm 0.3$. This value is consistent with the investigated
N-body simulation where a ratio of $2.5 \pm 0.2$ can be derived for
an edge-on galaxy with an edge-on bar.

\item 
The frequencies of bulge types in the simulated galaxy, seen under
different aspect angles of a bar and under high inclinations,
are consistent with the observed frequencies corrected 
for the total fraction of barred galaxies. 
These results support the
interpretation that bars viewed under different angles in edge-on
galaxies  play a major role for the bulge shape.
However, the viewing angle is not the only parameter, since
the size of the pre-existing spheroidal bulge and various bar shapes and types
also influence the bulge shape. 

\item Interpreting galaxies with
peanut bulges as galaxies with a bar seen edge-on
the ratio of bar length to thickness, ${\rm BAL/BAT}\!=\!14 \pm 4$,
can be directly measured for the first time.

\item
From the observational point of view bars and
b/p bulges are different components revealed by their
different lengths. 
The inner part of barred disk galaxies can be described by 
three components: the spheroidal bulge (including the central bulge),
the bar (=thin bar), and the b/p structure (=thick bar).

\item
We have found a  new size relation
of the positions of the maxima of the b/p distortion with the extent of the
central bulge (${\rm CBU/BPL}\!=\!1 \pm 0.1$). 
This size relation is most likely related to the in-plane ILR of the
bar. 

\end{itemize}

\begin{acknowledgements}
Part of this work was supported by the 
\emph{Deut\-sche For\-schungs\-ge\-mein\-schaft, DFG\/}.
Furthermore the authors want to thank Daniel Pfenniger who provided the
particle position file, Andreas Schr\"oer for helping with  IDL programming,
and the referee F. Combes for constructive comments.
\end{acknowledgements}

\end{document}